\documentclass[preprint,12pt,authoryear]{elsarticle}

\usepackage{amssymb}
\usepackage{amsmath}
\usepackage{graphicx}
\usepackage{enumerate}
\usepackage{natbib}
\usepackage{url} 
\usepackage{subfigure}
\usepackage{bm}
\usepackage{bbm}
\usepackage{amsmath,amssymb,amsfonts}

\newtheorem{remark}{Remark}[section]

\numberwithin{equation}{section}

\journal{Journal of Machine Learning Research}

\begin{document}

\begin{frontmatter}


  \title{Semi-supervised learning for linear extremile regression}
\author[mymainaddress]{Rong Jiang}
\author[mythirdaryaddress,myfourthaddress]{Keming Yu\corref{mycorrespondingauthor}}
\cortext[mycorrespondingauthor]{Corresponding author}
\ead{keming.yu@brunel.ac.uk}
\author[myfourthaddress]{Jiangfeng Wang}
\address[mymainaddress]{Shanghai University of International Business and Economics, People's Republic of China}
\address[mythirdaryaddress]{Brunel University, London UB83PH, UK}
\address[myfourthaddress]{Zhejiang Gongshang University, People's Republic of China}

\begin{abstract}
Extremile regression, as a least squares analog of quantile regression, is potentially useful tool for modeling and understanding the extreme tails of a distribution. However, existing extremile regression methods, as nonparametric approaches, may face challenges in high-dimensional settings due to data sparsity, computational inefficiency, and the risk of overfitting. While linear regression serves as the foundation for many other statistical and machine learning models due to its simplicity, interpretability, and relatively easy implementation, particularly  in high-dimensional settings, this paper introduces a novel definition of linear extremile regression along with an accompanying estimation methodology. The regression coefficient estimators of this method achieve $\sqrt{n}$-consistency, which nonparametric extremile regression may not provide. In particular, while semi-supervised learning can leverage unlabeled data to make more accurate predictions and avoid overfitting to small labeled datasets in high-dimensional spaces, we propose a semi-supervised learning approach to enhance estimation efficiency, even when the specified linear extremile regression model may be misspecified. Both simulation studies and real data analyses demonstrate the finite-sample performance of our proposed methods.
\end{abstract}

\end{frontmatter}



\section{Introduction}
Assessing the extreme behavior of random phenomena is a significant challenge across various fields, including finance, extreme weather and climate events, and medicine \citep{r74,r71,r78}. A common approach to addressing extreme events involves estimating the extreme quantile of a relevant random variable, such as the daily return of a stock market index or the intensity of an earthquake. At the same time, expectiles \citep{r70} can serve this purpose as well. While both quantiles and expectiles have been valuable tools, they have faced criticism in the literature for various axiomatic or practical reasons. Quantiles rely solely on whether an observation is below or above a specific threshold, whereas expectiles can lack transparent interpretation due to their lack of an explicit expression.
\par
Extremiles \citep{r1}, acting as the least squares analog of quantiles, offer a coherent risk measure based on weighted expectations, rather than relying on tail probabilities. Therefore, leveraging the definition of expected extreme value regression (which calculates the expectation of the minimum or maximum tail value) and its effectiveness in risk measurement, this approach can be employed for quantitative risk analysis of extreme events.

\cite{r41}  recently proposed
the conditional order-$\tau$ extremile of $\bm{Y}$ given $\bm{X}=\bm{x}$ as
\begin{equation}
	\begin{split}
		\xi_{\tau}(\bm{x})=\arg\min_{\theta\in \mathbb{R}}
		\text{E}\left[\text{J}_{\tau}\{\text{F}(\bm{Y|\bm{X}})\}\cdot(\bm{Y}-\theta)^2|\bm{X}=\bm{x}\right],
	\end{split}
\end{equation}
where $\text{F}(\cdot|\bm{X})$ is the conditional distribution of $\bm{Y}$ given $\bm{X}$, $\text{J}_{\tau}(t)=\partial\text{H}_{\tau}(t)/\partial t$ with
\begin{equation*}
	\begin{split}
		\text{H}_{\tau}(t)=\left \{
		\begin{array}{ll}
			1-(1-t)^{s(\tau)},&\textrm{if}~0< \tau\leq 1/2,\\
			t^{r(\tau)},&\textrm{if}~1/2\leq \tau<1,
		\end{array}
		\right.
	\end{split}
\end{equation*}
being a distribution function with support $[0,1]$ and
$r(\tau)=s(1-\tau)=\log(1/2)/\log(\tau)$.
\par
 Extremile is a novel and valuable concept, but there are currently few  studies on extremile. To date, only \cite{r50}, \cite{r80}, \cite{r81}, and \cite{r82} have addressed it, and none have introduced linear extremile regression. Whereas linear regression is one of the most widely used and popular regression techniques in statistics and data analysis. In particular, linear regression serves as the foundation for many other statistical and machine learning models due to its simplicity, interpretability, and relatively easy implementation, even with high-dimensional data. Therefore, we introduce the following linear extremile regression:
$
\xi_{\tau}(\bm{X})=\bm{X}^{\top}\bm{\beta}_{\tau},
$
then, from the equation (1.1) and \cite{r82}, we can obtain the estimator of $\bm{\beta}_{\tau}$ as:
\begin{equation}
	\begin{split}
		\bar{\bm{\beta}}_{\tau}=\arg\min_{\bm{\beta}\in \mathbb{R}} \sum_{i=1}^n\text{J}_{\tau}\{\hat{\text{F}}(Y_i|\bm{X}_i)\}\cdot(Y_i-\bm{X}_i^{\top}\bm{\beta})^2.
	\end{split}
\end{equation}

The estimator $\bar{\bm{\beta}}_{\tau}$   is difficult to achieve $\sqrt{n}$-consistency. For example, using the Nadaraya-Watson method to obtain $\hat{\text{F}}(Y_i|\bm{X}_i)$,  the convergence rate of $\bar{\bm{\beta}}_{\tau}$   with $\hat{\text{F}}(Y_i|\bm{X}_i)$ will be slower than $\sqrt{n}$, since $\hat{\text{F}}(Y_i|\bm{X}_i)$ is $\sqrt{nh}$  is $\sqrt{nh}$-consistent with $h\rightarrow 0$.
This poses a challenge for parameter estimators, as they are typically expected to be $\sqrt{nh}$-consistent. To address this issue, alternative parameter estimation methods for estimating $\text{F}(\bm{Y|\bm{X}})$
could be employed, although these often require strict conditions. We propose a new estimation method that avoids the inclusion of the unknown non-parametric component, as seen in equation (1.2). As a result, we are able to establish a $\text{F}(\bm{Y|\bm{X}})$-consistent estimator for the unknown parameters in the linear extremile regression model.

In particular,  with working on high-dimensional data, semi-supervised (SS) learning an be highly effective in domains where both the amount of labeled data is limited and the data itself is complex, with many features or variables, we present a methodology for utilizing unlabeled data to devise SS  learning methods for linear extremile regression. This approach not only constructs an effective estimation method when the working model is incorrectly specified but also performing as efficiently as supervised learning when the working model is correctly specified.

The SS setting encompasses two distinct datasets: (i) a labeled data set comprising observations for an outcome $\bm{Y}$ and a set of covariates $\bm{X}$, and (ii) a significantly larger unlabeled data set where only $\bm{X}$ is observed. This fundamental distinction sets SS settings apart from standard missing data problems, where the proportion is always assumed to be bounded away from 0, a condition commonly referred to as the positivity (or overlap) assumption in the missing data literature, which is inherently violated in this context.

For example, in biomedical applications, SS settings are assuming growing importance in modern integrative genomics, particularly in the study of expression quantitative trait loci (eQTL) \citep{r59}. This approach merges genetic association studies with gene expression profiles. However, a prevailing challenge in such studies is that due to the limited scale of costly gene expression data, their capabilities are often inadequate  \citep{r61}. Conversely, genetic variation recording is more cost-effective and can typically be used for large-scale datasets, naturally leading to SS settings. Additionally, SS settings are increasingly relevant in many fields like image processing \citep{r56}, anomaly detection \citep{r57}, empirical risk \citep{r55}. A comprehensive overview of SS learning and recent advancements can be found in \cite{r54} and \cite{r65}.  Related statistical theory research literature can be found in \citep{r64,r58,r62,r47,r79,r83}.
\par
To summarize, our statistical contributions are as follows:
\par
(i) In contract to the nonparametric setting of extremile regression \citep{r41}, we introduces a linear extremile regression for its simplicity, interpretability, and relatively easy implementation, particularly  in high-dimensional settings. For the estimation method, we estimate the quantile function (the inverse function of conditional distribution $\text{F}(\cdot|\bm{X})$)  instead of $\text{F}(\cdot|\bm{X})$.
In addition, we use a parametric method to estimate the quantile regression coefficients, which allows us to achieve $\sqrt{n}$-consistency for the unknown parameters in the linear extremile regression model.
\par
(ii) In cases of model misspecification, we have developed an estimation method for the unknown parameters in the linear extremile regression model using unlabeled data. We demonstrate that the resulting estimator is more effective than one that relies solely on labeled data. To the best of our knowledge, this is the first application of semi-supervised learning to extremile regression.
\par
The remaining sections of this paper are organized as follows: In Section 2, we introduce the new definition of linear extremile regression and its associated estimation method. Section 3 focuses on the development of semi-supervised learning. Section 4 presents simulation examples and applies real data to illustrate the proposed methods. Finally, we conclude the paper with a brief discussion in Section 5. All technical proofs are provided in the Appendix.

\section{The definition of a new linear extremile regression and its estimation method}
\subsection{The definition of linear extremile regression}
We define a new linear extremile regression and construct a $\sqrt{n}$-consistent estimator for $\bm{\beta}_{\tau}$ in the linear extremile regression.
Note that an alternative perspective on the $\tau$th extremile  of $\bm{Y}$ given $\bm{X}$ is the weighted quantile function (Proposition 1 in \cite{r41}):
\begin{equation}
	\begin{split}
		\xi_{\tau}(\bm{X})=\int_0^1q_{\bar{\tau}}(\bm{X})\text{J}_{\tau}(\bar{\tau})d\bar{\tau},
	\end{split}
\end{equation}
where $q_{\bar{\tau}}(\bm{X})$ is the conditional $\bar{\tau}$-th quantile of $\bm{Y}$ given $\bm{X}$.
Due to the linear assumption, only $q_{\bar{\tau}}(\bm{X})$ is related to $\bm{X}$ in the combination (2.1), thus it can be obtained
\begin{equation}
	\begin{split}
		q_{\bar{\tau}}(\bm{X})=\bm{X}^{\top}\bm{\gamma}(\bar{\tau}),
	\end{split}
\end{equation}
where $\bm{\gamma}(\cdot)$ is a $p$ dimensional unknown function.
Furthermore, we use the spline method to parameterize $\bm{\gamma}(\bar{\tau})$ as
\begin{equation}
	\begin{split}
		\bm{\gamma}(\bar{\tau})=\bm{\alpha}_0\bm{b}(\bar{\tau}),
	\end{split}
\end{equation}
where $\bm{b}(\bar{\tau})$ is the $q$ dimensional basis function and $\bm{\alpha}_0$ is a $p\times q$ unknown matrix.
Then, from (2.1)-(2.3), we can re-define the linear extremile regression as
\begin{equation}
	\begin{split}
		\xi_{\tau}(\bm{X})=		\bm{X}^{\top}\bm{\alpha}_0\int_0^1\bm{b}(\bar{\tau})\text{J}_{\tau}(\bar{\tau})d\bar{\tau}\equiv\bm{X}^{\top}\bm{\beta}_{\tau},
	\end{split}
\end{equation}
where
\begin{equation}
	\begin{split} \bm{\beta}_{\tau}=\bm{\alpha}_0\int_0^1\bm{b}(\bar{\tau})\text{J}_{\tau}(\bar{\tau})d\bar{\tau}.
	\end{split}
\end{equation}
\par
By comparing (1.2) and (2.5), we can see that only the estimation of unknown parameter $\bm{\alpha}_0$ is present in (2.5), which avoids a typically nonparametric  estimation of $\text{F}(\bm{Y|\bm{X}})$ in (1.2).
The following Proposition 2.1 confirms that the order-$\tau$ extremile $\xi_{\tau}(\bm{X})$ defined in (2.4) is the same as that in \cite{r41}.
\par
{\bf Proposition 2.1.}
Let $\bm{Y}$ given $\bm{X}$ have a finite absolute first moment. Then, for any $\tau\in (0,1)$, we have the following equivalent form expression:
\begin{equation*}
	\begin{split}
		\xi_{\tau}(\bm{X})=\left \{
		\begin{array}{ll}
			\text{E}\left\{\max(\bm{Y}_{\bm{X}}^1,\ldots,\bm{Y}_{\bm{X}}^{r})\right\},&\text{when}~\tau=0.5^{1/r}~\text{with}~r\in \mathbb{N}\backslash \{0\}\\
			\text{E}\left\{\min(\bm{Y}_{\bm{X}}^1,\ldots,\bm{Y}_{\bm{X}}^{s})\right\},&\text{when}~\tau=1-0.5^{1/s}~\text{with}~s\in \mathbb{N}\backslash\{0\},
		\end{array}
		\right.
	\end{split}
\end{equation*}
where $\{\bm{Y}_{\bm{X}}^i\}$ are independent observations and drawn from the conditional distribution of $\bm{Y}$ given $\bm{X}$.
\par
Specifically, when $\bm{X}=\bm{1}$, the linear extremile regression (2.4) is equal to extremile in \cite{r1} according to $\xi_{\tau}(\bm{X})=\bm{\beta}_{\tau}=\int_0^1\bm{\gamma}({\bar{\tau}})\text{J}_{\tau}(\bar{\tau})d\bar{\tau}=\int_0^1q_{\bar{\tau}}\text{J}_{\tau}(\bar{\tau})d\bar{\tau}=\xi_{\tau}$. Therefore, the proposed definition of linear extremile regression (2.4) is reasonable.

\subsection{Estimation method}
In this section, we estimate the unknown parameter $\bm{\beta}_{\tau}$ in (2.5) under dataset $\mathcal{L}=\{Y_i,{\bf X}_i\}_{i=1}^n$, which are $n$ independent and identically distributed (i.i.d) observations from $\{\bm{Y},\bm{X}^{\top}\}^{\top}$ in model (2.4).
\par
Based on (2.2) and (2.3), we have
$q_{\bar{\tau}}(\bm{X})=\bm{X}^{\top}\bm{\gamma}(\bar{\tau})=\bm{X}^{\top}\bm{\alpha}_0\bm{b}(\bar{\tau})$. Therefore, we can estimate
$\bm{\alpha}_0$ as the minimizer of the integrated objective function:
\begin{equation}
	\begin{split}
		\hat{\bm{\alpha}}=
		\arg\min_{\bm{\alpha}}\sum_{i=1}^{n}L(Y_i,\bm{X}_i,\bm{\alpha})
		=\arg\min_{\bm{\alpha}}\sum_{i=1}^{n}\int_0^1\rho_{\bar{\tau}}(Y_i-\bm{X}_i^{\top}\bm{\alpha}\bm{b}(\bar{\tau}))d\bar{\tau},
	\end{split}
\end{equation}
where  $L(Y_i,\bm{X}_i,\bm{\alpha})=\int_0^1\rho_{\bar{\tau}}(Y_i-\bm{X}_i^{\top}\bm{\alpha}\bm{b}(\bar{\tau}))d\bar{\tau}$ and $\rho_{\bar{\tau}}(r)=\bar{\tau}r-r\text{I}(r<0)$ is the quantile check function.
The objective function $L(Y_i,\bm{X}_i,\bm{\alpha})$ in (2.6)
can be regarded as an average loss function, achieved by marginalizing $
\rho_{\bar{\tau}}(\bm{Y}_i-\bm{X}_i^{\top}\bm{\alpha}\bm{b}(\bar{\tau}))$
over the entire interval $(0,1)$. In addition, the solution of minimizing (2.6) is currently implemented by the
\verb"iqr" function in the \verb"qrcm R" package.
Then, the estimator $\hat{\bm{\beta}}_{\tau}$ of $\bm{\beta}_{\tau}$ (2.5) with $\hat{\bm{\alpha}}$  (2.6) is
\begin{equation}
	\begin{split}
		\hat{\bm{\beta}}_{\tau}=
		\hat{\bm{\alpha}}\int_0^1\bm{b}(\bar{\tau})\text{J}_{\tau}(\bar{\tau})d\bar{\tau}.
	\end{split}
\end{equation}
\par
Note that (2.7), it permits estimating the entire extremile  process rather than only obtaining a discrete set of extremiles,
because $\hat{\bm{\alpha}}$
 is independent of $\bar{\tau}$.
This is an important property for streaming data analysis. Because, if a
quantile $\tau$ is not considered at the first batch, we cannot obtain its estimator in the subsequent batches due to the one-pass of streaming data \citep{r77}.
In addition, if $\int_0^1\bm{b}(\bar{\tau})\text{J}_{\tau}(\bar{\tau})d\bar{\tau}$ in (2.7) is not integrable, we can use
$n^{-1}\sum_{i=1}^n\bm{b}(i/n)\text{J}_{\tau}(i/n)$ to calculate it approximately.

\subsection{Large sample properties}
To facilitate the presentation of the derivation of the  asymptotic theories, let us introduce some notations.
Let $\text{Vec}(\cdot)$ be the vectoring operation, which creates a column vector by stacking the column vectors of below one another, that is,
$\text{Vec}(\bm{\alpha})=(\bm{\alpha}_1^{\top},\dots,\bm{\alpha}_q^{\top})^{\top}$
with $\bm{\alpha}_j=(\alpha_{1,j},\dots,\alpha_{p,j})^{\top}$.
Denote $S(\bm{\alpha})=n^{-1}\sum_{i=1}^{n}\nabla_{\text{Vec}(\bm{\alpha})}L(Y_i,\bm{X}_i,\bm{\alpha})
=n^{-1}\sum_{i=1}^{n}\int_0^1\bm{b}(\bar{\tau})\otimes\bm{X}_i[\text{I}(Y_i<\{\bm{b}(\bar{\tau})\otimes\bm{X}_i\}^{\top}\text{Vec}(\bm{\alpha}))-\bar{\tau}]d\bar{\tau}$,
where $\otimes$
is the Kronecker product.
\par
{\bf C1}: The true unknown parameter vector $\bm{\alpha}_0$ in (2.4) is an interior point of
a compact set $\Theta$ and
satisfies $E\left\{S(\bm{\alpha}_0)|\bm{X}\right\}={\bf 0}$.
\par
{\bf C2}: The loss function $L(\bm{Y},\bm{X},\bm{\alpha})$ satisfies $E[\sup_{\bm{\alpha}\in \Theta}\{L(\bm{Y},\bm{X},\bm{\alpha})\}^2]<\infty$. $S(\bm{\alpha})$ is continuously differentiable, $E\{\sup_{\bm{\alpha}\in \Theta}\|S(\bm{\alpha})\|_2^2\}<\infty$ and
\\
$E\{\sup_{\bm{\alpha}\in \Theta}\|\nabla_{\text{Vec}(\bm{\alpha})}S(\bm{\alpha})\|\}<\infty$
with $\|\cdot\|$ is the spectral norm.
\par
{\bf C3}: ${\bf H}=E\{\nabla_{\text{Vec}(\bm{\alpha})}S(\bm{\alpha})|_{\bm{\alpha}=\bm{\alpha}_0}\}=
E[\int_0^1\{\bm{X}^{\top}\bm{\alpha}_0\nabla_{\bar{\tau}}\bm{b}(\bar{\tau})\}^{-1}
\{\bm{b}(\bar{\tau})\otimes\bm{X}\}\{\bm{b}(\bar{\tau})\otimes\bm{X}\}^{\top}d\bar{\tau}]
$ is nonsingular.

\begin{remark}
The validity of the conditions {\bf C1} and {\bf C2} depends on the structure of
$\bm{b}(\bar{\tau})$, which should induce a well-defined quantile function  such that $\Theta$ is not empty (conditions {\bf C1}); $\bm{b}(\bar{\tau})$ is continuous
and ensure that a central limit theorem can be applied to $S(\bm{\alpha})$ (conditions {\bf C2}).	
Conditions {\bf C1} and {\bf C2} are also used in \cite{r44}. 	
Condition {\bf C3} is needed to establish the asymptotic normality.
\end{remark}
\par
{\bf Theorem 2.1.}
Suppose that the conditions {\bf C1} and {\bf C2} hold. Then as $n\rightarrow\infty$, we have
\begin{equation*}
\begin{split}
	\text{Vec}(\hat{\bm{\alpha}}-\bm{\alpha}_0)\xrightarrow{p} \bm{0},
\end{split}
\end{equation*}
where $\xrightarrow{p}$ represents the convergence in the probability.
Moreover, if condition {\bf C3} holds, we can obtain
\begin{equation*}
\begin{split}
	\sqrt{n}\text{Vec}(\hat{\bm{\alpha}}-\bm{\alpha}_0)
	\xrightarrow{L}\textrm{N}\left(\bm{0},{\bf H}^{-1}{\bm \Sigma}{\bf H}^{-1}\right),
\end{split}
\end{equation*}
where $\xrightarrow{L}$ represents the convergence in the distribution and ${\bm \Sigma}=E\{S(\bm{\alpha}_0)S(\bm{\alpha}_0)^{\top}\}$.
\\
\par
Theorem 2.1 shows that the parameter $\bm{\alpha}_0$ is identified and its estimator
$\hat{\bm{\alpha}}$ has a normal distribution in large samples. The large sample distribution of the plugin estimator of
$\bm{\beta}_{\tau}=\bm{\alpha}_0\int_0^1\bm{b}(\bar{\tau})\text{J}_{\tau}(\bar{\tau})d\bar{\tau}$ can also be obtained as
\begin{equation}
\begin{split}
	\hat{\bm{\beta}}_{\tau}= \hat{\bm{\alpha}}\int_0^1\bm{b}(\bar{\tau})\text{J}_{\tau}(\bar{\tau})d\bar{\tau}=\tilde{\bm{b}}(\tau)^{\top}\text{Vec}(\hat{\bm{\alpha}}),
\end{split}
\end{equation}
where $\tilde{\bm{b}}(\tau)=\int_0^1\bm{b}(\bar{\tau})\text{J}_{\tau}(\bar{\tau})d\bar{\tau}\otimes \bm{I}_p$ and $\bm{I}_p$ is the identity matrix of size $p$. Then, we consider the large sample distribution of $\hat{\bm{\beta}}_{\tau}$ by (2.8) in the following theorem.
\par
{\bf Theorem 2.2.}
Suppose that the conditions {\bf C1}-{\bf C3} hold and $\int_0^1\bm{b}(\bar{\tau})\text{J}_{\tau}(\bar{\tau})d\bar{\tau}$ is finite, we have
\begin{equation*}
\begin{split}
	\sqrt{n}(\hat{\bm{\beta}}_{\tau}-\bm{\beta}_{\tau})
	\xrightarrow{L}\textrm{N}\left(\bm{0},\tilde{\bm{b}}(\tau)^{\top}{\bf H}^{-1}{\bm \Sigma}{\bf H}^{-1}\tilde{\bm{b}}(\tau)\right).
\end{split}
\end{equation*}

\section{Semi-supervised learning}

\subsection{Data representation}
Let $\mathbb{P}$ be the joint distribution of $\{\bm{Y},\bm{X}^{\top}\}^{\top}$ and let $\mathbb{P}_{X}$ be the marginal distribution of $\bm{X}$.
In semi-supervised setting, the data available are $\mathcal{D}=\mathcal{L}\cup \mathcal{M}$,
where $\mathcal{L}=\{Y_i,{\bf X}_i\}_{i=1}^n$ is from $\mathbb{P}$ and $\mathcal{M}=\{{\bf X}_i\}_{i=n+1}^N$ with $N\geq 1$ are $N$ i.i.d observations from $\mathbb{P}_{X}$. The $n/N\rightarrow\rho$ for some constant $\rho\in[0,+\infty)$ as $n\rightarrow\infty$ and $N\rightarrow\infty$. Note that the
semi-supervised setting allows $n/N\rightarrow0$, that means that the unlabeled dataset can be of much larger size than the labeled one in various practical problems, as labeling of the outcomes is often very costly. However, the missing completely at random assumption is that $\rho>0$, which is the major difference between semi-supervised setting and missing data.
\subsection{The target parameter}
In most real-world data analyses, the model (2.4) may not be correct because the linear structure assumptions of the model are too strong.
But due to the simplicity and interpretability of the linear structure, it is often continued to be used. Therefore, consider a $\tau$th linear extremile working regression model
$\xi_{\tau}(\bm{X})=\bm{X}^{\top}\bm{\alpha}^{*}\int_0^1\bm{b}(\bar{\tau})\text{J}_{\tau}(\bar{\tau})d\bar{\tau}$,
where the unknown parameter $\bm{\alpha}^{*}$ is defined as
\begin{equation}
\begin{split}
	\bm{\alpha}^{*}=&\arg\min_{\bm{\alpha}}E\{L(\bm{Y},\bm{X},\bm{\alpha})\}.
\end{split}
\end{equation}
\par
It is noteworthy that in supervised framework (section 2.2), $\bm{\alpha}^{*}$ is equal to $\bm{\alpha}_0$ when the outcome variable $\bm{Y}$ is fully observed and the working model is correctly specified.
\subsection{Semi-supervised learning}
In order to construct a consistent estimator of $E\{L(\bm{Y},\bm{X},\bm{\alpha})\}$ based on the unlabeled data,
we select a random vector $\bm{Z}\in \mathbb{R}^d$, which is a function of $\bm{X}$ for some positive integer $d$ such that
\begin{equation*}
\begin{split}
	E\{L(\bm{Y},\bm{X},\bm{\alpha})\}=E\{\bm{Z}^{\top}\varphi(\bm{\alpha})\},
\end{split}
\end{equation*}
where $\varphi(\bm{\alpha})=\{E(\bm{Z}\bm{Z}^{\top})\}^{-1}E\{\bm{Z}L(\bm{Y},\bm{X},\bm{\alpha})\}$.
Without loss of generality, we fix the first element of $\bm{Z}$ to be 1. Different $\bm{Z}$ provides us with different ways to incorporate the
unlabeled data, such as $\bm{Z}=(1,\bm{X}^{\top},\ldots,(\bm{X}^d)^{\top})^{\top}$.
The Selection of $\bm{Z}$ can see section 3.2 in \cite{r47}.
Then, we propose a new class of loss functions by incorporating the
information from the unlabeled data into the supervised estimation:
\begin{equation}
\begin{split}
	\tilde{\bm{\alpha}}=&\arg\min_{\bm{\alpha}}\left\{
	\sum_{i=1}^{n}L(Y_i,\bm{X}_i,\bm{\alpha})
	+\sum_{i=n+1}^{n+N}\bm{Z}_i^{\top}\hat{\varphi}(\bm{\alpha})\right\}\\
	=&\arg\min_{\bm{\alpha}}\sum_{i=1}^{n}\omega_iL(Y_i,\bm{X}_i,\bm{\alpha}),
\end{split}
\end{equation}
where
$\hat{\varphi}(\bm{\alpha})=\left(\sum_{i=1}^{n}\bm{Z}_i\bm{Z}_i^{\top}/n\right)^{-1}
\sum_{i=1}^{n}\bm{Z}_iL(Y_i,\bm{X}_i,\bm{\alpha})/n$
and
$$\omega_i=1+N/n\left(\sum_{i=n+1}^{n+N}\bm{Z}_i/N\right)^{\top}
\left(\sum_{i=1}^{n}\bm{Z}_i\bm{Z}_i^{\top}/n\right)^{-1}\bm{Z}_i.$$
\par
In addition, the solution of minimizing (3.2) is currently implemented by the
\verb"iqr" function in the \verb"qrcm R" package. From the equations (2.5) and (3.2), we can obtain
\begin{equation}
\begin{split}
	\tilde{\bm{\beta}}_{\tau}=
	\tilde{\bm{\alpha}}\int_0^1\bm{b}(\bar{\tau})\text{J}_{\tau}(\bar{\tau})d\bar{\tau}.
\end{split}
\end{equation}

\subsection{Large sample properties}
The following conditions are needed to establish the consistency and asymptotic normality of $\tilde{\bm{\alpha}}$.
\par
{\bf C4}: The unknown parameter vector $\bm{\alpha}^{*}$ in (3.1) is an interior point of
a compact set $\Theta$,
and $\tilde{{\bf H}}=E\{\nabla_{\text{Vec}(\bm{\alpha})}S(\bm{\alpha})|_{\bm{\alpha}=\bm{\alpha}^{*}}\}
$ is nonsingular.
\par
{\bf C5}: The random vector $\bm{Z}$ is bounded almost surely and ${\bm \Sigma}_{\bm{Z}}=E(\bm{Z}\bm{Z}^{\top})$
is nonsingular.
\begin{remark}
The condition {\bf C4} ensures the uniqueness of $\bm{\alpha}^{*}$ under the strict convexity of $E\{L(\bm{Y},\bm{X},\bm{\alpha})\}$.
Condition {\bf C5} is regularity condition for the unlabeled data. Conditions {\bf C4} and {\bf C5} are also used in \cite{r47}.
\end{remark}
\par
{\bf Theorem 3.1.}
Suppose that the conditions {\bf C2}, {\bf C4} and {\bf C5} hold and $n/N\rightarrow\rho$ for some constant $\rho\in[0,+\infty)$. Then as $n\rightarrow\infty$ and $N\rightarrow\infty$, we have
\begin{equation*}
\begin{split}
	\text{Vec}(\tilde{\bm{\alpha}}-\bm{\alpha}^{*})\xrightarrow{p} \bm{0},
\end{split}
\end{equation*}
and
\begin{equation*}
\begin{split}
	\sqrt{n}\text{Vec}(\tilde{\bm{\alpha}}-\bm{\alpha}^{*})
	\xrightarrow{L}\textrm{N}\left(\bm{0},\tilde{{\bf H}}^{-1}{\bm \Sigma}_{\rho}\tilde{{\bf H}}^{-1}\right),
\end{split}
\end{equation*}
where ${\bm \Sigma}_{\rho}=E(\bm{W}\bm{W}^{\top})+\rho E(\bm{V}\bm{V}^{\top})$,
$\bm{W}=S(\bm{\alpha}^{*})-N(n+N)^{-1}\bm{A}^{\top}\bm{Z}$,
$\bm{V}=N(n+N)^{-1}\bm{A}^{\top}\bm{Z}$ and $\bm{A}={\bm \Sigma}_{\bm{Z}}^{-1}E\{\bm{Z}S(\bm{\alpha}^{*})^{\top}\}$.
\\
\par
{\bf Theorem 3.2.}
Suppose that conditions in Theorem 3.1 hold and $\int_0^1\bm{b}(\bar{\tau})\text{J}_{\tau}(\bar{\tau})d\bar{\tau}$ is finite, we have
\begin{equation*}
\begin{split}
	\sqrt{n}(\tilde{\bm{\beta}}_{\tau}-\bm{\beta}^{*}_{\tau})
	\xrightarrow{L}\textrm{N}\left(\bm{0},\tilde{\bm{b}}(\tau)^{\top}\tilde{{\bf H}}^{-1}{\bm \Sigma}_{\rho}\tilde{{\bf H}}^{-1}\tilde{\bm{b}}(\tau)\right),
\end{split}
\end{equation*}
where $\tilde{\beta}_{\tau}=\tilde{\alpha}\int_0^1\bm{b}(\bar{\tau})\text{J}_{\tau}(\bar{\tau})d\bar{\tau}$ and
$\bm{\beta}^{*}_{\tau}=\bm{\alpha}^{*}\int_0^1\bm{b}(\bar{\tau})\text{J}_{\tau}(\bar{\tau})d\bar{\tau}$.

\subsection{Comparison between supervised learning and semi-supervised learning}
Theorems 2.2 and 3.2 have important implication on the asymptotic efficiency comparison between the supervised estimator $\hat{\bm{\beta}}_{\tau}$ (2.7) and the semi-supervised estimator $  \tilde{\bm{\beta}}_{\tau}$ (3.3).
From Theorem 2.2, ${\bm \Sigma}$ can be written as the following form as
$\bm{U}$ and $\bm{A}^{\top}\bm{Z}$ are uncorrelated,
$${\bm \Sigma}=E(\bm{U}\bm{U}^{\top})+E\left\{(\bm{A}^{\top}\bm{Z})
(\bm{A}^{\top}\bm{Z})^{\top}\right\},$$ where $\bm{U}=S(\bm{\alpha}^{*} )-\bm{A}^{\top}\bm{Z}$. Moreover, ${\bm \Sigma}_{\rho}$
can be rewrite as:
$${\bm \Sigma}_{\rho}=E(\bm{U}\bm{U}^{\top})+\frac{n}{n+N}E\left\{(\bm{A}^{\top}\bm{Z})(\bm{A}^{\top}\bm{Z})^{\top}\right\}.$$
Note that $n/(n+N)\leq 1$.
Therefore, the semi-supervised estimator $\tilde{\bm{\beta}}_{\tau}$
is equally or more efficient than the supervised estimator $\hat{\bm{\beta}}_{\tau}$
according to ${\bm \Sigma}_{\rho}\leq{\bm \Sigma}$.
\subsection{Estimation of covariance}
Finally, we provide consistent analytical estimators of the components of the variances below:
\begin{equation*}
\begin{split}
	&\hat{{\bf H}}(\bm{\alpha})=\frac{1}{n}\sum_{i=1}^{n}
	\int_0^1\{\bm{X}_i^{\top}\bm{\alpha}\nabla_{\bar{\tau}}\bm{b}(\bar{\tau})\}^{-1}
	\{\bm{b}(\bar{\tau})\otimes\bm{X}_i\}\{\bm{b}(\bar{\tau})\otimes\bm{X}_i\}^{\top}d\bar{\tau},\\
	&\hat{{\bm \Sigma}}_{\rho}= \frac{1}{n}\sum_{i=1}^{n}\hat{\bm{W}}_i\hat{\bm{W}}_i^{\top}+\frac{1}{N}\sum_{i=n+1}^{n+N}\hat{\bm{V}}_i\hat{\bm{V}}_i^{\top},\\
	&\hat{{\bm \Sigma}}=\frac{1}{n}\sum_{i=1}^{n}\hat{S}_i(\hat{\bm{\alpha}})\hat{S}_i(\hat{\bm{\alpha}})^{\top},
\end{split}
\end{equation*}
where $\hat{\bm{W}}_i=\hat{S}_i(\tilde{\bm{\alpha}})-N(n+N)^{-1}\hat{\bm{A}}^{\top}\bm{Z}_i,$
$\hat{\bm{V}}_i=N(n+N)^{-1}\hat{\bm{A}}^{\top}\bm{Z}_i,$
$\hat{S}_i(\bm{\alpha})=\bm{b}(\bar{\tau})\otimes\bm{X}_i\int_0^1[\text{I}(Y_i<\{\bm{b}(\bar{\tau})\otimes\bm{X}_i\}^{\top}\text{Vec}(\bm{\alpha}))-\bar{\tau}]d\bar{\tau}$
and
$\hat{\bm{A}}=(\sum_{i=1}^{n}\bm{Z}_i\bm{Z}_i)^{-1}\sum_{i=1}^{n}\{\bm{Z}_i\hat{S}_i(\tilde{\bm{\alpha}})^{\top}\}$.
The consistency of these estimators follows from the law of large numbers and consistency of $\text{Vec}(\hat{\bm{\alpha}})$ and $\text{Vec}(\tilde{\bm{\alpha}})$, as all the components are continuous functions of the parameters. Therefore, the limiting covariance matrices $\tilde{\bm{b}}(\tau)^{\top}{\bf H}^{-1}{\bm \Sigma}{\bf H}^{-1}\tilde{\bm{b}}(\tau)$ and $\tilde{\bm{b}}(\tau)^{\top}\tilde{{\bf H}}^{-1}{\bm \Sigma}_{\rho}\tilde{{\bf H}}^{-1}\tilde{\bm{b}}(\tau)$ in Theorems 2.2 and 3.2
can be estimated by $\tilde{\bm{b}}(\tau)^{\top}\{\hat{{\bf H}}(\hat{\bm{\alpha}})\}^{-1}\hat{{\bm \Sigma}}\{\hat{{\bf H}}(\hat{\bm{\alpha}})\}^{-1}\tilde{\bm{b}}(\tau)$ and $\tilde{\bm{b}}(\tau)^{\top}\{\hat{{\bf H}}(\tilde{\bm{\alpha}})\}^{-1}\hat{{\bm \Sigma}}_{\rho}\{\hat{{\bf H}}(\tilde{\bm{\alpha}})\}^{-1}\tilde{\bm{b}}(\tau)$, respectively.

\section{Numerical studies}
In this section, we first use Monte Carlo simulation studies to assess the finite sample performance of the proposed procedures and then demonstrate the application of the proposed methods with three real data analysis.
We choose $\bm{b}(\tau)=(\tau,3/2\tau^2-1/2,5/2\tau^3-3/2\tau)$ in this section, which is a 3rd-degree shifted Legendre polynomial and the
default setting in the \verb"iqr" function in the \verb"qrcm R" package.
All programs are written in \textsf{R} code.
\subsection{Simulation example 1: the performance of new linear extremile regression}
In this section, we study the performances of the new linear extremile regression in section 2. Specifically, we compare the supervised learning (SL) $\hat{\bm{\beta}}_{\tau}$ with ordinary estimator (OE) in (1.2) by using Nadaraya-Watson method to estimate $\text{F}(\cdot|{\bm X})$.
We generate data from the following linear model:
\begin{equation}
\begin{split}
	{\bm Y}={\bm X}^{\top}\bm{\beta}_0+\sigma({\bm X})(\bm{\varepsilon}-\hat{e}_{\tau}),
\end{split}
\end{equation}
where ${\bm X}=({\bm 1},{\bm X}_{1},{\bm X}_{2})^{\top}$, ${\bm X}_{1}$ and ${\bm X}_{2}$ are drawn from a uniform distribution $U(0,1)$.
The true value of the parameter is $\bm{\beta}_0=(1,2,3)^{\top}$.
The $
\hat{e}_{\tau}=\sum_{i=1}^{\tilde{n}}
\left\{\text{H}_{\tau}\left(\frac{i}{\tilde{n}}\right)-\text{H}_{\tau}\left(\frac{i-1}{\tilde{n}}\right)\right\}\varepsilon_{i,\tilde{n}}
$ is the estimator of the $\tau$-level extremile of $\bm{\varepsilon}$ \citep{r1}, which is used to make the true value of ${\bm\beta}_0$ always $(1,2,3)^{\top}$ under different $\tau$ values,
where $\varepsilon_{1,\tilde{n}}\leq \varepsilon_{2,\tilde{n}}\leq\cdots\leq \varepsilon_{\tilde{n},\tilde{n}}$ denotes the ordered sample and $\tilde{n}=10^6$.
Therefore, we have $\xi_{\tau}(\bm{X})=\bm{X}^{\top}\bm{\beta}_{0}$ by the setting of model (4.1).
Three error distributions of $\bm{\varepsilon}$ are considered:
a standard normal distribution $N(0,1)$, a t distribution with 5 degrees of freedom $t(5)$ and a uniform distribution $U(0,1)$. Two cases of $\sigma({\bf X})$ are considered:
$\sigma({\bf X})=0.5$ and $\sigma({\bf X})=0.4\sqrt{1+|{\bf X}_{1}|+|{\bf X}_{2}|}$.
\par
To evaluate the performance of the estimation method, we calculate the total absolute error : $\text{TAE}=\sum_{j=1}^3|\hat{\bm{\beta}}_{\tau,j}-\bm{\beta}_{0,j}|$ and the percentage of relative TAE by OE and SL as $\text{PRTAE}=(\text{TAE}_{OE}-\text{TAE}_{SL})/\text{TAE}_{SL}\times 100\%$.
The simulation results of the means of TAEs and PRTAEs based on $\tau=0.1,0.3,0.5,0.7,0.9$ and sample size $n=500$ are
shown in Tables 4.1 and 4.2, which are based on 500 simulation replications.
The simulation results in Tables 4.1 and 4.2 show that the performances of SL are better than those of OE under different errors, $\tau$s and $\sigma({\bf X})$. This means that our proposed method SL has indeed improved the estimation accuracy compared to OE.
This is also consistent with the theoretical results. Our proposed estimate is $\sqrt{n}$ consistent, while OE is lower than $\sqrt{n}$ consistent.
In addition, it can be observed that at the extreme quantile level (where extremile regression focuses), our results (SL) are significantly better than those of OE. Although SL and OE are very close when $\tau=0.5$, for extremile regression, we do not pay as much attention to  $\tau=0.5$ as quantile regression does.
\begin{table}[htp]
\footnotesize
\caption{The means and standard deviations (in parentheses) of TAEs with different errors, $\tau$s and methods under $\sigma({\bf X})=0.5$.  }
\centering
\begin{tabular}{@{}c|c|ccccc@{}}
	\hline
	Error&Method&$\tau=0.1$&$\tau=0.3$&$\tau=0.5$&$\tau=0.7$&$\tau=0.9$\\
	\hline
	N(0,1)&$\text{TAE}_{OE}$& 0.526~(0.220)& 0.295~(0.135)& 0.179~(0.093)& 0.224~(0.112)& 0.525~(0.232)\\
	&$\text{TAE}_{SL}$& 0.250~(0.131)& 0.188~(0.101)& 0.178~(0.092)& 0.188~(0.098)& 0.259~(0.136)\\
	&PRTAE& 110.4\%& 56.9\%& 0.6\%& 19.2\%& 102.7\%\\
	\hline
	t(5)&$\text{TAE}_{OE}$& 0.759~(0.369)& 0.340~(0.169)& 0.230~(0.120)& 0.295~(0.158)& 0.808~(0.388)\\
	&$\text{TAE}_{SL}$& 0.384~(0.179)& 0.237~(0.120)& 0.222~(0.113)& 0.242~(0.125)& 0.388~(0.184)\\
	&PRTAE& 97.7\%& 43.5\%& 3.6\%& 21.9\%& 108.2\%\\
	\hline
	U(0,1)&$\text{TAE}_{OE}$& 0.203~(0.079)& 0.113~(0.039)& 0.051~(0.026)& 0.080~(0.020)& 0.131~(0.055)\\
	&$\text{TAE}_{SL}$& 0.044~(0.034)& 0.051~(0.028)& 0.051~(0.026)& 0.052~(0.028)& 0.042~(0.023)\\
	&PRTAE& 361.4\%& 121.6\%& 0.0\%& 53.8\%& 211.9\%\\
	\hline
\end{tabular}
\end{table}

\begin{table}[htp]
\footnotesize
\caption{The means and standard deviations (in parentheses) of TAEs with different errors, $\tau$s and methods under $\sigma({\bf X})=0.4\sqrt{1+|{\bf X}_{1}|+|{\bf X}_{2}|}$.  }
\centering
\begin{tabular}{@{}c|c|ccccc@{}}
	\hline
	Error&Method&$\tau=0.1$&$\tau=0.3$&$\tau=0.5$&$\tau=0.7$&$\tau=0.9$\\
	\hline
	N(0,1)&$\text{TAE}_{OE}$& 0.520~(0.221)& 0.330~(0.155)& 0.182~(0.094)& 0.243~(0.128)& 0.513~(0.256)\\
	&$\text{TAE}_{SL}$& 0.275~(0.140)& 0.199~(0.106)& 0.181~(0.093)& 0.202~(0.104)& 0.284~(0.143)\\
	&PRTAE& 89.1\%& 65.8\%& 0.6\%& 20.3\%& 80.6\%\\
	\hline
	t(5)&$\text{TAE}_{OE}$& 0.808~(0.442)& 0.371~(0.194)& 0.249~(0.134)& 0.308~(0.163)& 0.815~(0.427)\\
	&$\text{TAE}_{SL}$& 0.413~(0.203)& 0.244~(0.133)& 0.234~(0.124)& 0.264~(0.130)& 0.426~(0.203)\\
	&PRTAE& 95.6\%& 52.0\%& 6.4\%& 16.7\%& 91.3\%\\
	\hline
	U(0,1)&$\text{TAE}_{OE}$& 0.195~(0.065)& 0.112~(0.039)& 0.056~(0.031)& 0.089~(0.022)& 0.152~(0.073)\\
	&$\text{TAE}_{SL}$& 0.047~(0.030)& 0.056~(0.030)& 0.055~(0.031)& 0.056~(0.028)& 0.044~(0.022)\\
	&PRTAE& 314.9\%& 100.0\%& 1.8\%& 58.9\%& 245.5\%\\
	\hline
\end{tabular}
\end{table}

\subsection{Simulation example 2: the performance of semi-supervised learning}
In this section, we study the performances of the semi-supervised learning in section 3.
Therefore, we generate data based on the following nonlinear regression model in \cite{r47}:
\begin{equation}
\begin{split}
	{\bf Y}=\alpha_0+{\bf X}^{\top}\bm{\alpha}_1+\alpha_2\sum_{j,k}{\bm X}_{j}{\bm X}_{k}+(1+{\bm X}^{\top}\bm{\alpha}_3)\bm{\varepsilon},
\end{split}
\end{equation}
where ${\bm X}=({\bm X}_{1},\dots,{\bm X}_{4})^{\top}$, $\{{\bm X}_{j}\}_{j=1}^4$ are drawn from standard normal distribution, and the true value of the parameter is $(\alpha_0,\bm{\alpha}_1^{\top},\alpha_2,\bm{\alpha}_3^{\top})^{\top}=(1,0.5,0.5,0.5$, $0.5,1,0.5,0.5,0,0)^{\top}$. Three error distributions of $\bm{\varepsilon}$ are considered:
$N(0,1)$, $t(5)$ and $U(0,1)$. The labeled $(n)$ and three unlabeled $(N)$ sample size are considered:
$n=500$ and $N=500,1000,2000$.
\par
Due to the data being generated from the nonlinear model (4.2), the linear extremile regression model of the working model is incorrect. At this time, we will examine the effect of semi-supervised learning $\tilde{\bm{\beta}}_{\tau}$ (SSL) in (3.3).
Moreover, we compare SSL with the supervised learning (SL) in (2.7) by the
estimated percentage asymptotic relative efficiency (PARE), which is defined by the ratio of the empirical variance (EVar) of estimator of SL and that of SSL. Specifically defined as $\text{PARE}=(\text{EVar}_{SL}-\text{EVar}_{SSL})/\text{EVar}_{SSL}\times 100\%$.
The simulation results of the means of estimators, their empirical standard deviations and PAREs based on $\tau=0.1,0.3,0.5,0.7,0.9$ are shown in Tables 4.3-4.5, which are based on 500 simulation replications.
From the results in Tables 4.3-4.5, we can see that the improvements of semi-supervised learning (SSL)
over the supervised learning (SL) are more significant when $N$ gets large according to PARE, which is reasonable.
In addition, from Figure 4.1, it can be seen that in most cases (82.2\%), PARE is greater than 20\%, a few parts (4.9\%) are less than 10\%, and there is a 34.7\% proportion of PARE greater than 50\%. This indicates that semi-supervised learning (SSL) is significant more efficient than supervised learning (SL) under model mis-specification.

\begin{table}[htp]
\footnotesize
\caption{The means, standard deviations (in parentheses) and PAREs of SL and SSL with different $\tau$s and Ns under $\bm{\varepsilon}\sim N(0,1)$.  }
\centering
\begin{tabular}{@{}c|c|ccc|rrr@{}}
	\hline
	&&&SSL&&&PARE\\
	\cline{3-8}
	$\tau$&SL&N=500&N=1000&N=2000&N=500&N=1000&N=2000\\
	\hline
	0.1 & 1.772~(0.115) & 1.783~(0.112)& 1.785~(0.110)& 1.786~(0.109)& ~5.3\%&~8.8\%&12.1\%\\
	& 0.054~(0.197) & 0.058~(0.193)& 0.059~(0.188)& 0.057~(0.187)& ~5.1\%&~9.6\%&10.9\%\\
	& 0.070~(0.195) & 0.073~(0.181)& 0.070~(0.180)& 0.073~(0.174)& 15.3\%&17.1\%&25.7\%\\
	& 0.491~(0.196) & 0.492~(0.186)& 0.491~(0.185)& 0.493~(0.176)& 10.5\%&11.3\%&23.4\%\\
	& 0.508~(0.212) & 0.505~(0.200)& 0.508~(0.192)& 0.502~(0.188)& 12.8\%&21.7\%&27.0\%\\
	\hline
	0.3 & 3.420~(0.109) & 3.436~(0.094)& 3.438~(0.087)& 3.439~(0.084)& 34.1\%& 55.3\%&66.2\%\\
	& 0.304~(0.159) & 0.297~(0.140)& 0.301~(0.138)& 0.295~(0.132)& 28.6\%& 32.7\%&43.5\%\\
	& 0.295~(0.152) & 0.300~(0.138)& 0.292~(0.132)& 0.295~(0.130)& 21.9\%& 32.9\%&37.6\%\\
	& 0.507~(0.155) & 0.502~(0.139)& 0.504~(0.136)& 0.503~(0.125)& 24.5\%& 30.2\%&54.8\%\\
	& 0.501~(0.161) & 0.504~(0.149)& 0.507~(0.138)& 0.507~(0.133)& 17.1\%& 35.7\%&45.7\%\\
	\hline
	0.5 & 4.907~(0.133) & 4.925~(0.102)& 4.931~(0.092)& 4.931~(0.084)& 71.6\%&111.7\%&153.1\%\\
	& 0.391~(0.162) & 0.392~(0.137)& 0.395~(0.136)& 0.393~(0.128)& 39.7\%& 41.8\%&59.5\%\\
	& 0.393~(0.158) & 0.393~(0.138)& 0.393~(0.134)& 0.398~(0.127)& 29.7\%& 38.8\%&53.0\%\\
	& 0.502~(0.157) & 0.503~(0.134)& 0.503~(0.133)& 0.501~(0.124)& 36.7\%& 39.2\%&58.8\%\\
	& 0.501~(0.152) & 0.497~(0.135)& 0.496~(0.124)& 0.501~(0.122)& 27.2\%& 52.0\%&56.3\%\\
	\hline
	0.7 & 6.419~(0.187) & 6.427~(0.150)& 6.425~(0.126)& 6.430~(0.109)& 55.5\%&119.3\%&192.8\%\\
	& 0.508~(0.185) & 0.515~(0.160)& 0.510~(0.150)& 0.512~(0.141)& 33.2\%& 50.6\%&70.8\%\\
	& 0.499~(0.190) & 0.509~(0.164)& 0.507~(0.151)& 0.507~(0.139)& 34.6\%& 59.5\%&86.0\%\\
	& 0.496~(0.174) & 0.496~(0.150)& 0.490~(0.147)& 0.493~(0.138)& 34.3\%& 40.2\%&58.5\%\\
	& 0.502~(0.195) & 0.501~(0.160)& 0.510~(0.151)& 0.500~(0.144)& 47.6\%& 65.4\%&82.0\%\\
	\hline
	0.9 & 8.929~(0.295) & 8.945~(0.260)& 8.957~(0.255)& 8.954~(0.251)& 28.8\%& 33.9\%&38.1\%\\
	& 0.581~(0.252) & 0.587~(0.227)& 0.582~(0.212)& 0.586~(0.203)& 23.2\%& 40.9\%&54.0\%\\
	& 0.607~(0.260) & 0.605~(0.241)& 0.605~(0.234)& 0.600~(0.229)& 16.2\%& 22.8\%&28.4\%\\
	& 0.504~(0.244) & 0.508~(0.220)& 0.500~(0.218)& 0.501~(0.214)& 22.1\%& 25.2\%&29.7\%\\
	& 0.514~(0.260) & 0.510~(0.238)& 0.511~(0.214)& 0.513~(0.206)& 19.2\%& 47.6\%&58.8\%\\
	\hline
\end{tabular}
\end{table}

\begin{table}[htp]
\footnotesize
\caption{The means, standard deviations (in parentheses) and PAREs of SL and SSL with different $\tau$s and Ns under $\bm{\varepsilon}\sim t(5)$.  }
\centering
\begin{tabular}{@{}c|c|ccc|rrr@{}}
	\hline
	&&&SSL&&&PARE\\
	\cline{3-8}
	$\tau$&SL&N=500&N=1000&N=2000&N=500&N=1000&N=2000\\
	\hline
	0.1 &~1.611~(0.129) &~1.625~(0.125)&~1.631~(0.125)&~1.634~(0.123)& 6.1\%& 6.5\%& 10.0\%\\
	&-0.064~(0.222) &-0.070~(0.205)&-0.067~(0.204)&-0.074~(0.201)&17.3\%& 18.4\%& 21.9\%\\
	&-0.090~(0.210) &-0.083~(0.204)&-0.085~(0.199)&-0.081~(0.198)& 6.5\%& 12.1\%& 12.6\%\\
	&~0.498~(0.216) &~0.498~(0.209)&~0.493~(0.203)&~0.494~(0.202)& 6.3\%& 12.6\%& 14.3\%\\
	&~0.503~(0.222) &~0.500~(0.212)&~0.502~(0.206)&~0.500~(0.204)& 9.8\%& 16.2\%& 19.5\%\\
	\hline
	0.3 & 3.370~(0.117) & 3.378~(0.102)& 3.383~(0.094)& 3.385~(0.090)& 30.6\%& 56.3\%& 70.6\%\\
	& 0.222~(0.157) & 0.218~(0.140)& 0.216~(0.139)& 0.219~(0.137)& 25.9\%& 27.2\%& 30.2\%\\
	& 0.229~(0.168) & 0.225~(0.147)& 0.223~(0.143)& 0.221~(0.136)& 31.8\%& 39.2\%& 54.3\%\\
	& 0.494~(0.174) & 0.493~(0.152)& 0.493~(0.151)& 0.494~(0.144)& 31.8\%& 31.3\%& 46.2\%\\
	& 0.501~(0.174) & 0.504~(0.160)& 0.501~(0.149)& 0.505~(0.143)& 17.5\%& 36.0\%& 48.1\%\\
	\hline
	0.5 &~4.922~(0.140) &~4.926~(0.109)&~4.932~(0.102)&~4.937~(0.091)& 63.6\%& 89.1\%& 137.2\%\\
	&~0.362~(0.169) &~0.367~(0.145)&~0.373~(0.132)&~0.367~(0.131)& 36.0\%& 64.6\%& 66.2\%\\
	&~0.369~(0.169) &~0.364~(0.152)&~0.366~(0.137)&~0.370~(0.132)& 22.5\%& 51.3\%& 63.3\%\\
	&~0.508~(0.171) &~0.504~(0.151)&~0.508~(0.137)&~0.505~(0.133)& 27.9\%& 56.5\%& 64.4\%\\
	&~0.499~(0.164) &~0.499~(0.144)&~0.499~(0.137)&~0.499~(0.129)& 29.5\%& 44.0\%& 60.5\%\\
	\hline
	0.7 & 6.474~(0.193) & 6.483~(0.152)& 6.485~(0.139)& 6.485~(0.122)& 61.5\%& 92.4\%&150.3\%\\
	& 0.520~(0.201) & 0.518~(0.166)& 0.518~(0.160)& 0.517~(0.153)& 45.7\%& 56.3\%& 71.9\%\\
	& 0.511~(0.188) & 0.513~(0.155)& 0.515~(0.153)& 0.510~(0.146)& 46.2\%& 51.0\%& 64.9\%\\
	& 0.515~(0.187) & 0.512~(0.160)& 0.510~(0.148)& 0.507~(0.144)& 36.7\%& 60.0\%& 69.3\%\\
	& 0.516~(0.190) & 0.512~(0.162)& 0.513~(0.152)& 0.510~(0.140)& 37.6\%& 56.3\%& 83.3\%\\
	\hline
	0.9 & 9.069~(0.294) & 9.099~(0.270)& 9.092~(0.267)& 9.096~(0.244)& 18.2\%& 21.0\%& 45.7\%\\
	& 0.646~(0.250) & 0.639~(0.245)& 0.645~(0.236)& 0.648~(0.219)& 4.2\%& 12.1\%& 30.0\%\\
	& 0.649~(0.270) & 0.658~(0.245)& 0.657~(0.231)& 0.654~(0.224)& 21.7\%& 37.4\%& 45.9\%\\
	& 0.512~(0.249) & 0.509~(0.228)& 0.502~(0.221)& 0.508~(0.210)& 19.2\%& 27.2\%& 41.1\%\\
	& 0.513~(0.253) & 0.503~(0.235)& 0.511~(0.223)& 0.510~(0.218)& 16.2\%& 28.8\%& 34.6\%\\
	\hline
\end{tabular}
\end{table}

\begin{table}[htp]
\footnotesize
\caption{The means, standard deviations (in parentheses) and PAREs of SL and SSL with different $\tau$s and Ns under $\bm{\varepsilon}\sim U(0,1)$.  }
\centering
\begin{tabular}{@{}c|c|ccc|rrr@{}}
	\hline
	&&&SSL&&&PARE\\
	\cline{3-8}
	$\tau$&SL&N=500&N=1000&N=2000&N=500&N=1000&N=2000\\
	\hline
	0.1 &2.640~(0.074) &2.626~(0.067)&2.622~(0.067)&2.616~(0.066)& 23.7\%& 24.5\%& 25.9\%\\
	&0.695~(0.147) &0.696~(0.134)&0.692~(0.130)&0.693~(0.129)& 21.0\%& 28.1\%& 30.0\%\\
	&0.690~(0.158) &0.694~(0.145)&0.691~(0.141)&0.696~(0.133)& 18.4\%& 26.6\%& 42.6\%\\
	&0.504~(0.157) &0.505~(0.146)&0.503~(0.145)&0.506~(0.143)& 15.6\%& 17.9\%& 20.8\%\\
	&0.507~(0.156) &0.504~(0.143)&0.506~(0.138)&0.504~(0.137)& 17.9\%& 27.5\%& 29.0\%\\
	\hline
	0.3 & 4.046~(0.098) & 4.042~(0.074)& 4.036~(0.067)& 4.038~(0.055)& 79.3\%& 113.2\%& 218.3\%\\
	& 0.720~(0.134) & 0.724~(0.115)& 0.720~(0.110)& 0.725~(0.106)& 35.7\%& 50.1\%& 59.8\%\\
	& 0.719~(0.142) & 0.716~(0.120)& 0.717~(0.110)& 0.719~(0.099)& 39.2\%& 64.6\%& 103.3\%\\
	& 0.494~(0.136) & 0.494~(0.124)& 0.494~(0.113)& 0.494~(0.107)& 20.1\%& 44.2\%& 60.8\%\\
	& 0.515~(0.143) & 0.506~(0.127)& 0.512~(0.115)& 0.512~(0.108)& 28.1\%& 54.0\%& 76.9\%\\
	\hline
	0.5 &~5.409~(0.129) &~5.419~(0.095)&~5.430~(0.078)&~5.436~(0.063)& 85.8\%& 177.6\%& 323.1\%\\
	&~0.730~(0.144) &~0.731~(0.125)&~0.735~(0.117)&~0.731~(0.112)& 33.2\%& 51.8\%& 65.9\%\\
	&~0.737~(0.152) &~0.735~(0.127)&~0.737~(0.121)&~0.735~(0.107)& 43.8\%& 57.8\%& 103.1\%\\
	&~0.502~(0.147) &~0.505~(0.129)&~0.502~(0.115)&~0.501~(0.110)& 29.0\%& 64.4\%& 78.0\%\\
	&~0.495~(0.156) &~0.494~(0.130)&~0.500~(0.118)&~0.498~(0.105)& 44.5\%& 73.4\%& 122.0\%\\
	\hline
	0.7 & 6.759~(0.186) & 6.784~(0.134)& 6.793~(0.120)& 6.806~(0.099)& 92.4\%& 141.5\%& 251.2\%\\
	& 0.749~(0.172) & 0.747~(0.142)& 0.751~(0.139)& 0.747~(0.128)& 47.1\%& 52.8\%& 79.3\%\\
	& 0.764~(0.167) & 0.759~(0.142)& 0.763~(0.132)& 0.758~(0.126)& 39.0\%& 61.3\%& 76.6\%\\
	& 0.497~(0.165) & 0.504~(0.146)& 0.504~(0.138)& 0.500~(0.127)& 27.2\%& 43.0\%& 68.2\%\\
	& 0.495~(0.167) & 0.495~(0.138)& 0.500~(0.127)& 0.492~(0.127)& 46.7\%& 74.2\%& 73.2\%\\
	\hline
	0.9 & 9.148~(0.284) & 9.183~(0.254)& 9.185~(0.238)& 9.181~(0.236)& 25.4\%& 42.6\%& 45.0\%\\
	& 0.765~(0.236) & 0.759~(0.222)& 0.768~(0.209)& 0.759~(0.200)& 13.6\%& 28.6\%& 39.9\%\\
	& 0.772~(0.237) & 0.773~(0.211)& 0.766~(0.209)& 0.775~(0.201)& 25.7\%& 28.6\%& 38.5\%\\
	& 0.497~(0.254) & 0.498~(0.230)& 0.506~(0.223)& 0.494~(0.215)& 22.1\%& 29.5\%& 39.5\%\\
	& 0.512~(0.222) & 0.502~(0.203)& 0.502~(0.196)& 0.496~(0.184)& 19.5\%& 28.8\%& 46.2\%\\
	\hline
\end{tabular}
\end{table}

\begin{figure}
\centering
\includegraphics[width=4in]{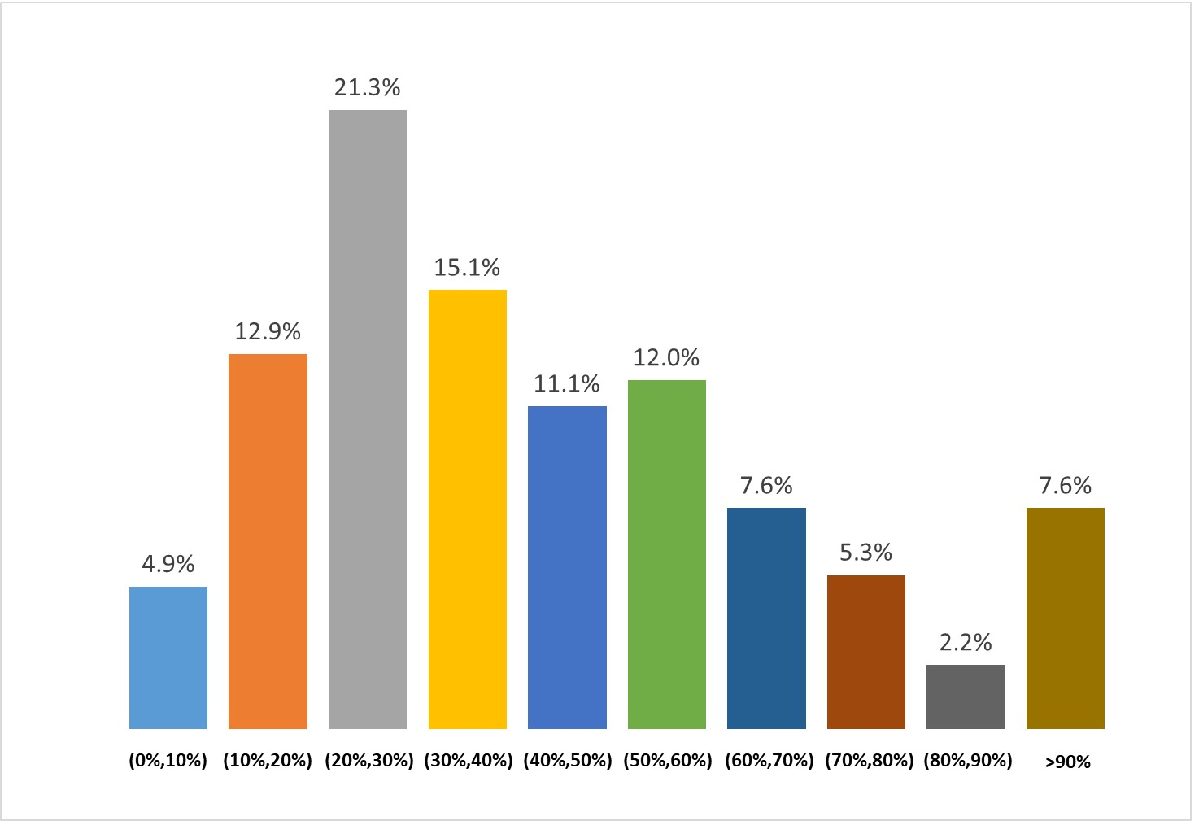}
\caption{The percentage of PARE in the Tables 4.3-4.5.}
\end{figure}

\subsection{Real data application 1: the motorcycle insurance data}
To illustrate the proposed linear extremile regression in section 2, we analyzed motorcycle insurance data, which can be obtained from the dataset ``dataOhlsson'' in the R package ``insuranceData''.
The data comes from Wasa, a former Swedish insurance company, which involved partial casco insurance for motorcycles. It contains 670 motorcycle-related claims recorded from 1994 to 1998.
\par
In this study, we investigate the linear relationship between claim costs (in thousands of US dollars) and owner age (0 to 99 years old). From scatter plot in Figure 4.2, it can be observed that there are multiple outliers present, therefore quantiles, expectiles and extremiles are more suitable for analyzing this dataset.
Moreover, \cite{r41} has also studied this dataset by extremiles, which is based on a non-parametric model.
The linear fits of quantiles, expectiles and extremiles under quantile levels 0.05, 0.1, 0.3, 0.5, 0.6, 0.7, 0.75, 0.8, 0.85, 0.9, 0.95 are given in Figure 4.2.
There are two instances of crossover in the quantile curves plot: the 0.7 and 0.75 quantile curves and the 0.9 and 0.95 quantile curves. Extremile and expectile curves do not exhibit this phenomenon.
In addition, the extremiles seem to afford a middle course between quantiles and expectiles. Through this example analysis, we can find that extremiles does not have the unreasonable crossover phenomenon that often occurs in quantiles, and the effect is between quantiles and expectiles.

\begin{figure}
\centering
\includegraphics[width=6in]{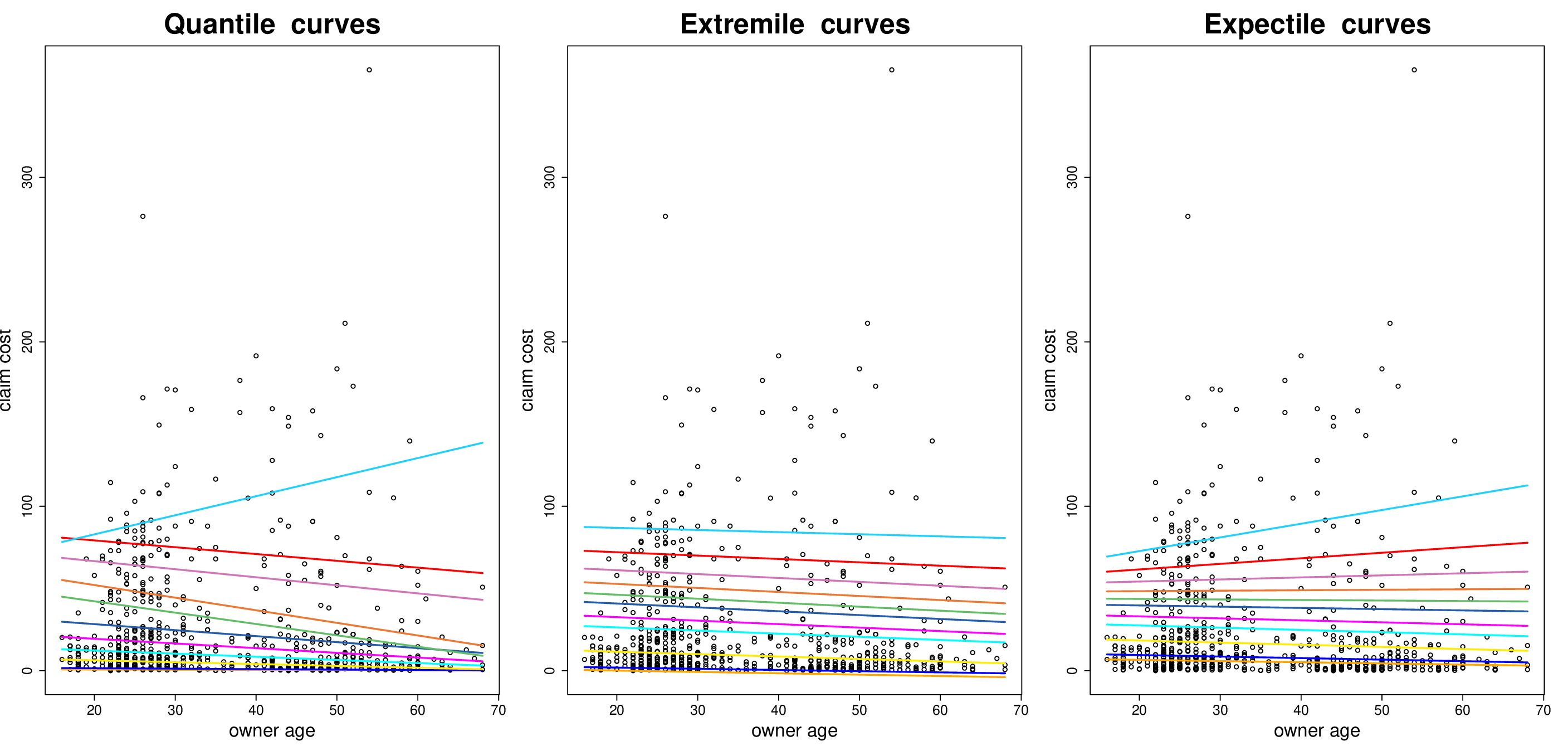}
\caption{The quantile (left), extremile (middle) and expectile (right) curves of quantile levels 0.05, 0.1, 0.3, 0.5, 0.6, 0.7, 0.75, 0.8, 0.85, 0.9, 0.95.}
\end{figure}

\subsection{Real data application 2: the mass body index (BMI) data}
We compare the proposed supervised learning (SL) in (2.7) with ordinary estimator (OE) in (1.2) using BMI dataset. The dataset include 2111 records for the estimation of obesity levels in individuals from the countries of Mexico, Peru and Colombia.
We use the $\text{BMI}=\text{weight(kg)}/\text{height}^2(\text{m})$ to measure obesity. The detailed description of the dataset can be found in \cite{r73}. The dataset can be downloaded from the following website: https://archive.ics.uci.edu/dataset/544/estimation+of+obesity+
levels+based+on+eating+habits+and+physical+condition.
\par
In this study, we investigated the linear relationship between BMI and gender (0 for female and 1 for male), age (14 to 61 years), physical activity frequency (PAF) and time using
technology devices (TUE) by quantiles, expectiles and extremiles (based on SL and OE) according to the left heavy tail phenomenon in Figure 4.3.
The estimated coefficients by quantile, expectile, extremile-SL and extremile-OE under quantile levels from 0.05 to 0.95 with a step size of 0.05 are given in Figure 4.4.
From Figure 4.4, it can be observed that our proposed supervised learning (extremile-SL)
conforms to the above phenomenon (between quantile and expectil), while the ordinary estimator (extremile-OE) clearly does not conform to the situation of Age and TUE. Moreover, the estimated coefficients of quantiles fluctuate greatly, while extremile-SLs and expectils are relatively smooth, and extremile-OEs also fluctuate greatly under Age and TUE. Therefore, the proposed supervised learning (SL) in (2.7) is better than the ordinary estimator (OE) in (1.2).

\begin{figure}
\centering
\includegraphics[width=5in]{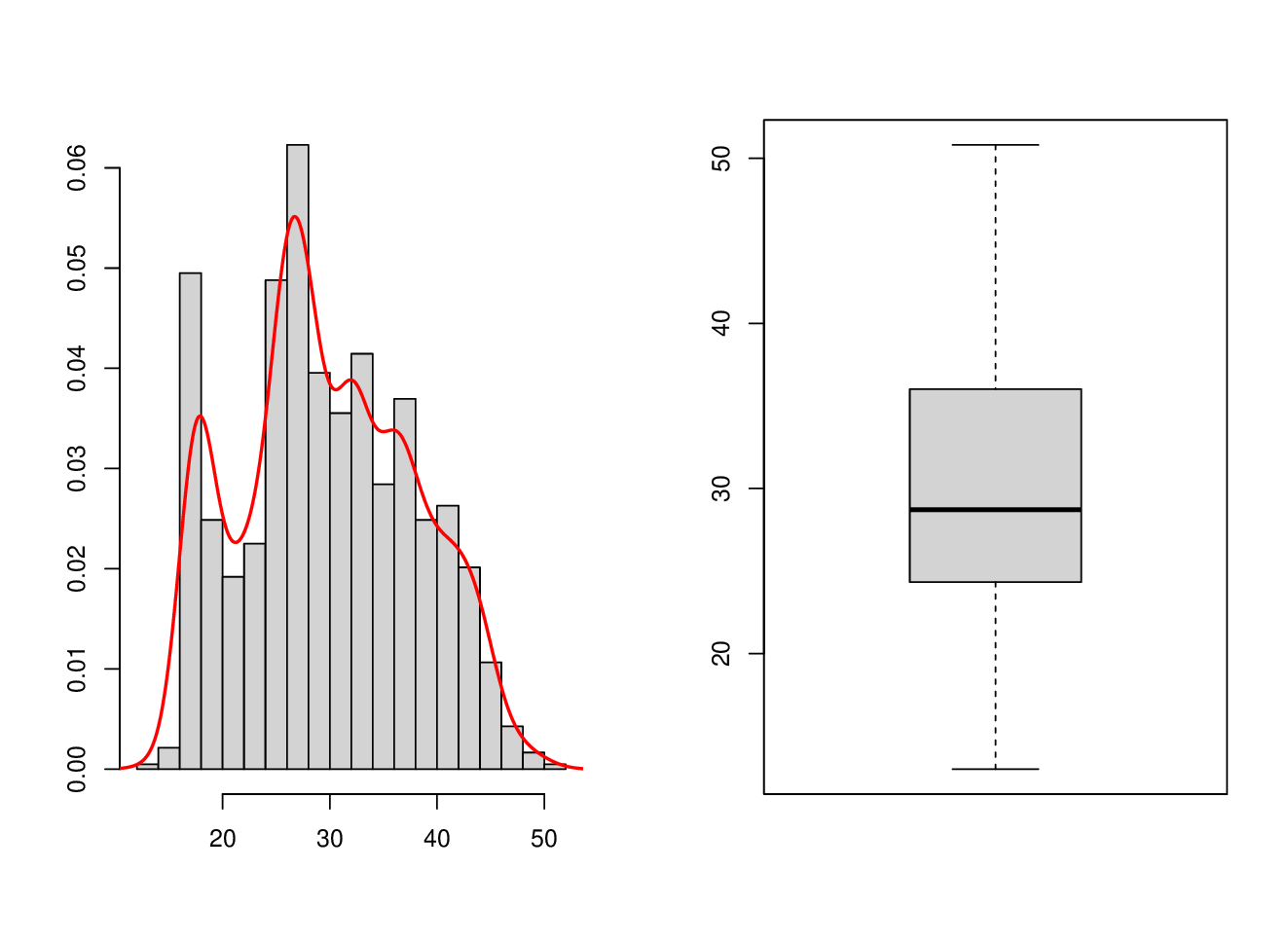}
\caption{Histogram and Box plot of BMI.}
\end{figure}

\begin{figure}
\centering
\includegraphics[width=5in]{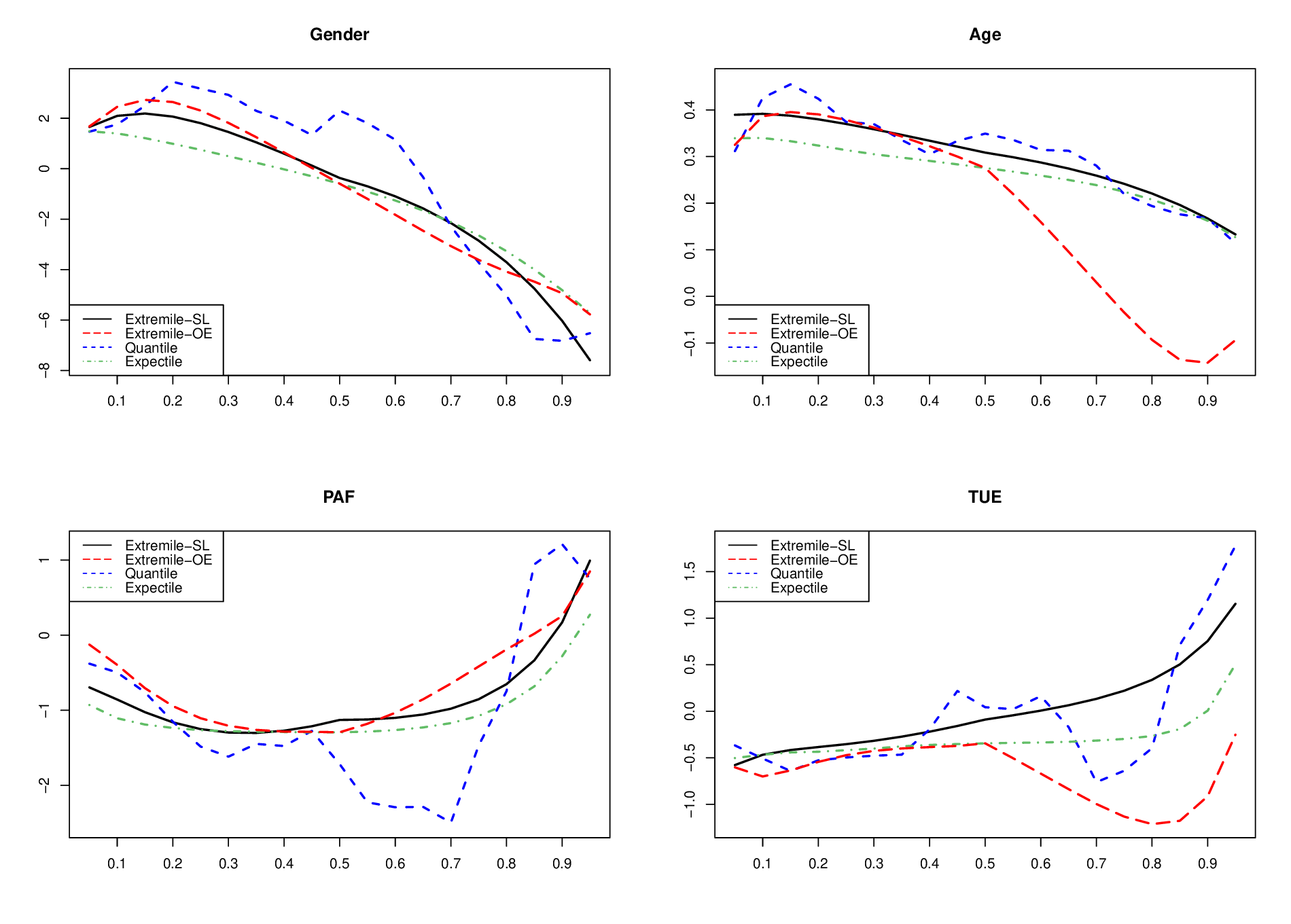}
\caption{The estimated coefficients (vertical axis) by quantile, expectile, extremile-SL and extremile-OE under quantile levels (horizontal axis) from 0.05 to 0.95 with a step size of 0.05.}
\end{figure}

\subsection{Real data application 3: the homeless data in Los Angeles County}
To illustrate the proposed semi-supervised learning in section 3, we used a data set, which are the total number of people estimated by the Los Angeles Homeless Services Administration (LAHSA) to be on the streets, shelters, or ``almost homeless'' in Los Angeles County from 2004 to 2005.
Because the entire County of Los Angeles includes 2,054 census tracts, it is expensive to survey the entire county, so stratified spatial sampling of census tracts is used.
It calls for two steps.
In the first step, they visited areas believed to have large numbers of homeless people, known as ``hot tracts''. The second step is to randomly draw stratified samples from the population in non-hot areas. Stratified random sampling 265 tracts, the remaining 1545 tracts were not visited.
Therefore, the dataset with total 1810 observations (labeled data 265 and unlabeled data 1545) is considered. This dataset can be downloaded in the supplemental material of \cite{r47} (https://www.tandfonline.com/doi/suppl/10.1080/01621459.2023.2169699?scroll=top).
\par
From the histogram and box plot of the homeless in Figure 4.5, there are multiple large outliers. Therefore, we use the linear extremile regression model (2.4) to analyze the change in the homeless street count over the effect of four predictors, which are
PctVacant (\% of unoccupied housing units), PctOwnerOcc (\% of owner-occupied housing units), PctMinority (\% of population that is non-Caucasian) and
MHI (Median household income).
They are the important variable in \cite{r52}. The estimators of regression coefficients, theirs estimated standard deviations (estimated by the methods in section 3.6) and PAREs (defined in section 4.2)
are present in Table 4.6. The results show that the proposed semi-supervised estimators generally have significant improvement over the supervised estimators based on the values of PARE being positive (except for the Intercept term, which is not of concern).
Moreover, in most cases, the increase is greater than 10\%, and there is even an increase of 44.1\%. Therefore, semi-supervised learning is more effective for this dataset.

\begin{figure}
\centering
\includegraphics[width=4in]{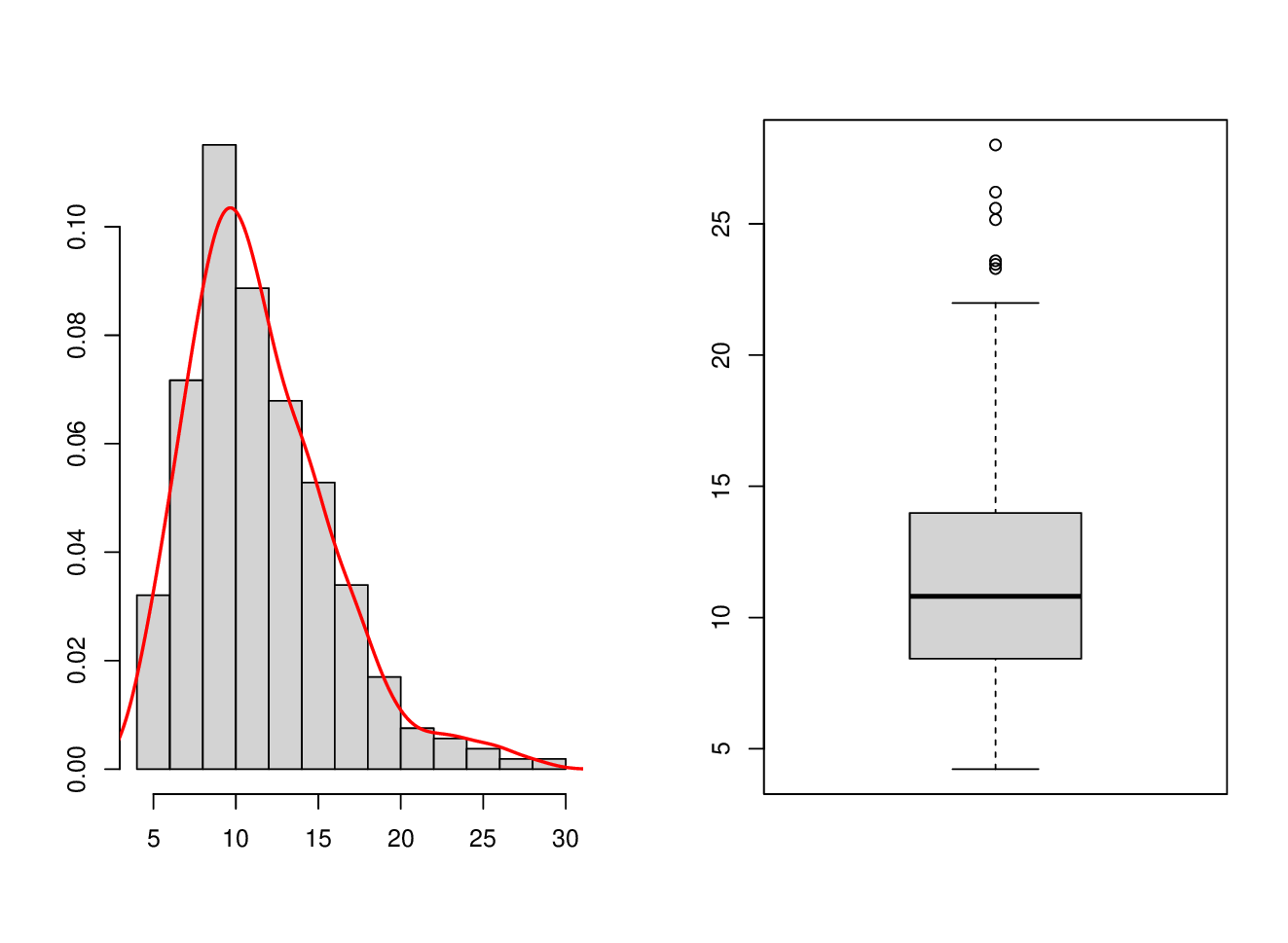}
\caption{Histogram and Box plot of the homeless.}
\end{figure}

\begin{table}[htp]
\footnotesize
\caption{The estimators and theirs estimated standard deviations (Esd) of SL and SSL and PAREs with different $\tau$s for the homeless data in Los Angeles County.  }
\centering
\begin{tabular}{@{}c|c|cccccccccc@{}}
	\hline
	$\tau$&Method&Intercept&PctVacant&PctOwnerOcc&PctMinority&MHI\\
	\hline
	0.1 &Estimator of SL &6.908 &1.027&0.895&0.824&0.713\\
	&Estimator of SSL&7.213 &0.962&0.886&0.745&1.015\\
	&Esd of SL        &0.430 &0.394&0.430&0.412&0.376\\
	&Esd of SSL       &0.457 &0.368&0.416&0.375&0.350\\
	&PARE             &-11.4\% &15.1\%&6.7\%&21.2\%&15.9\%\\
	\hline
	0.3 &Estimator of SL &9.286 &0.757&0.811&0.515&0.833\\
	&Estimator of SSL&9.688 &0.794&0.914&0.472&1.207\\
	&Esd of SL        &0.558 &0.508&0.533&0.539&0.533\\
	&Esd of SSL       &0.572 &0.476&0.511&0.496&0.488\\
	&PARE             &-4.9\% &13.8\%&8.9\%&18.2\%&19.5\%\\
	\hline
	0.5 &Estimator of SL &11.352 &0.710&0.813&0.326&0.887\\
	&Estimator of SSL&11.815 &0.780&1.055&0.239&1.151\\
	&Esd of SL        &~0.608 &0.558&0.567&0.559&0.609\\
	&Esd of SSL       &~0.626 &0.522&0.543&0.520&0.539\\
	&PARE             &-5.6\% &14.5\%&9.2\%&15.6\%&27.3\%\\
	\hline
	0.7 &Estimator of SL &13.407 &0.659&0.812&0.136&0.943\\
	&Estimator of SSL&13.933 &0.763&1.193&0.008&1.101\\
	&Esd of SL        &~0.674 &0.620&0.609&0.587&0.701\\
	&Esd of SSL       &~0.694 &0.579&0.582&0.552&0.606\\
	&PARE             &-5.5\% &14.9\%&9.4\%&13.0\%&33.9\%\\
	\hline
	0.9 &Estimator of SL &16.703 &0.790&0.867&-0.089&0.983\\
	&Estimator of SSL&17.298 &0.915&1.488&-0.443&0.825\\
	&Esd of SL        &~0.752 &0.680&0.664&~0.581&0.810\\
	&Esd of SSL       &~0.776 &0.624&0.636&~0.552&0.675\\
	&PARE             &-6.2\% &18.8\%&9.0\%&10.9\%&44.1\%\\
	\hline
\end{tabular}
\end{table}
\section{Conclusion}
The article introduces a novel approach to linear extremile regression that does not rely on estimating an unknown distribution function, as in \cite{r1,r41}. Instead, it achieves $\sqrt{n}$-consistent estimators for the unknown regression parameters, which is of significant theoretical importance and aligns with expectations in parametric regression analysis. Furthermore, the proposed estimation methods can easily accommodate various
?$\tau$-extremiles, making them well-suited for big data analysis. In particular, we explore semi-supervised settings and, through Theorems 2.2 and 3.2 and simulation studies, demonstrate the effectiveness of estimates obtained from both labeled  and unlabeled data.
\par
Based on the findings of this article, it becomes feasible to extend the application of "extremile" to more complex models, such as single-index models \citep{r66} and varying coefficient models \citep{r69}, as well as to complex data types, including massive data \citep{r68} and streaming data \citep{r76}.
\section*{Acknowledgments}
The research is supported by the National Social Science Foundation of China (Grant No. 21BTU040 and  20BTJ049), National Natural Science Foundation of China (Grant No. U23A2064),
Zhejiang Provincial Natural Science Foundation (Grant No. LY24A010004).

\appendix
\section{Proof of Theorems} 
{\bf Proof of Proposition 2.1.} Based on (2.1)-(2.5), we can obtain that
\begin{equation*}
	\begin{split}
		\xi_{\tau}(\bm{X})&=\bm{X}^{\top}\bm{\beta}_{\tau}=
		\bm{X}^{\top}\int_0^1\bm{\gamma}(\bar{\tau})\text{J}_{\tau}
		(\bar{\tau})d\bar{\tau}=\int_0^1\bm{X}^{\top}\bm{\gamma}(\bar{\tau})
		\text{J}_{\tau}(\bar{\tau})d\bar{\tau}\\
		&=\int_0^1q_{\bar{\tau}}(\bm{X})\text{J}_{\tau}(\bar{\tau})d\bar{\tau}
		=\int_{y\in \mathbb{R}}y\text{J}_{\tau}\{\text{F}(y|\bm{X})\}d\text{F}(y|\bm{X})\\
		&=E[\bm{Y}\text{J}_{\tau}\{\text{F}(\bm{Y}|\bm{X})\}]=E(\bm{Z}_{\bm{X},{\tau}}),
	\end{split}
\end{equation*}
where $\bm{Z}_{\bm{X},{\tau}}$ has cumulative distribution function $\text{F}_{\bm{Z}_{\bm{X},{\tau}}}=\text{H}_{\tau}\{\text{F}(\cdot|\bm{X})\}$.
When $\tau=0.5^{1/r}$ and $r\in \mathbb{N}\backslash \{0\}$, for any $z\in \mathbb{R}$, we have
\begin{equation*}
	\begin{split}
		\text{F}_{\bm{Z}_{\bm{X},{\tau}}}(z)=\text{H}_{\tau}\{\text{F}(z|\bm{X})\}
=\{\text{F}(z|\bm{X})\}^r=P\left(\max(Y_{\bm{X}}^1,\ldots,Y_{\bm{X}}^{r})\leqslant z\right),
	\end{split}
\end{equation*}
so that $\xi_{\tau}(\bm{X})
=E(\bm{Z}_{\bm{X},{\tau}})=E\{\max(Y_{\bm{X}}^1,\ldots,Y_{\bm{X}}^{r})\}$ according to
$\bm{Z}_{\bm{X},{\tau}}=\max(Y_{\bm{X}}^1,\ldots,Y_{\bm{X}}^{r})$. Similarly, we can prove that $\xi_{\tau}(\bm{X})
=E\{\min(Y_{\bm{X}}^1,\ldots,Y_{\bm{X}}^{r})\}$ under $\tau=1-0.5^{1/s}$ and $s\in \mathbb{N}\backslash \{0\}$. Because,
for any $z\in \mathbb{R}$,
\begin{equation*}
	\begin{split} 1-\text{F}_{\bm{Z}_{\bm{X},{\tau}}}(z)=1-\text{H}_{\tau}\{\text{F}(z|\bm{X})\}=\{1-\text{F}(z|\bm{X})\}^s
		=P\left(\min(Y_{\bm{X}}^1,\ldots,Y_{\bm{X}}^{r})> z\right).
	\end{split}
\end{equation*}
\\
\par
{\bf Proof of Theorems 2.1 and 2.2.} The results of Theorems 2.1 and 2.2 can be obtained directly from the following proofs of theorems 3.1 and 3.2 under $N=0$ and $\bm{\alpha}^*=\bm{\alpha}_0$.
\\
\par
{\bf Proof of Theorem 3.1.}
{\bf To establish consistency.}
Denote
\begin{equation*}
	\begin{split}
		\tilde{L}(\bm{\alpha})=\sum_{i=1}^{n}L(Y_i,\bm{X}_i,\bm{\alpha})
		+\sum_{i=n+1}^{n+N}\bm{Z}_i^{\top}\hat{\varphi}(\bm{\alpha}).
	\end{split}
\end{equation*}
Note that $\tilde{L}(\bm{\alpha})$ is the loss function in equation (3.2). We first show that
$\tilde{L}(\bm{\alpha})$ is invariant under affine transformation on $\bm{Z}=(1,\tilde{\bm{Z}}^{\top})^{\top}$, where $\tilde{\bm{Z}}$ is the remainder of $\bm{Z}$ after removing the first element. Assume that $\tilde{\bm{Z}}_i=\bm{M}\tilde{\bm{Q}}_i+\bm{b}$ for any fixed nonsingular $(d-1)\times (d-1)$ matrix $\bm{M}$ and $d-1$ vector $\bm{b}$, where $\tilde{\bm{Z}}_i$ is the $i$-th observation of $\tilde{\bm{Z}}$. Then, we can rewrite $\bm{Z}_i$ as
$$
\bm{Z}_i=\left(
\begin{array}{ll}
	1&{\bf 0}^{\top}_{q-1}\\
	\bm{b}&\bm{M}
\end{array}
\right)
\bm{Q}_i,
$$
where $\bm{Q}_i=(1,\tilde{\bm{Q}}_i^{\top})^{\top}$. Then, for any $\bm{\alpha}$, we can obtain
\begin{equation*}
	\begin{split}
		\tilde{L}(\bm{\alpha})=&
		\sum_{i=1}^{n}L(Y_i,\bm{X}_i,\bm{\alpha})
		+\sum_{i=n+1}^{n+N}\bm{Z}_i^{\top}
		\left(\frac{1}{n}\sum_{i=1}^{n}\bm{Z}_i\bm{Z}_i^{\top}\right)^{-1}
		\frac{1}{n}\sum_{i=1}^{n}\bm{Z}_iL(Y_i,\bm{X}_i,\bm{\alpha})\\
		=&\sum_{i=1}^{n}L(Y_i,\bm{X}_i,\bm{\alpha})
		+\sum_{i=n+1}^{n+N}\bm{Q}_i^{\top}
		\left(\frac{1}{n}\sum_{i=1}^{n}\bm{Q}_i\bm{Q}_i^{\top}\right)^{-1}
		\frac{1}{n}\sum_{i=1}^{n}\bm{Q}_iL(Y_i,\bm{X}_i,\bm{\alpha}).
	\end{split}
\end{equation*}
Therefore, we consider $E(\bm{Z}\bm{Z}^{\top})=\bm{I}_d$ and $E(\bm{Z})=(1,\bm{0}_{d-1}^{\top})^{\top}$ in the following proofs according to the above affine transformation invariant property.
Let
\begin{equation}
	\begin{split}
		\bar{L}(\bm{\alpha})=&\sum_{i=1}^{n}L(Y_i,\bm{X}_i,\bm{\alpha})
		+NE(\bm{Z}^{\top})\{E(\bm{Z}\bm{Z}^{\top})\}^{-1}
		\frac{1}{n}\sum_{i=1}^{n}\bm{Z}_iL(Y_i,\bm{X}_i,\bm{\alpha})\\
		=&\frac{n+N}{n}\sum_{i=1}^{n}L(Y_i,\bm{X}_i,\bm{\alpha}).
	\end{split}
\end{equation}
\par
Next, we proof that $\sup_{\bm{\alpha}\in\Theta}|\tilde{L}(\bm{\alpha})-\bar{L}(\bm{\alpha})|=o_p(1)$. Then, by Lemma 1 in \cite{r48} and conditions {\bf C2} and {\bf C5}, for large enough constants $c_1$ and $c_2$, we have
\begin{equation}
	\begin{split}
		&|\tilde{L}(\bm{\alpha})-\bar{L}(\bm{\alpha})|\\
		=&
		\left|\left[\sum_{i=n+1}^{n+N}\bm{Z}_i^{\top}
		\left(\frac{1}{n}\sum_{i=1}^{n}\bm{Z}_i\bm{Z}_i^{\top}\right)^{-1}
		-NE(\bm{Z}^{\top})\{E(\bm{Z}\bm{Z}^{\top})\}^{-1}\right]
		\frac{1}{n}\sum_{i=1}^{n}\bm{Z}_iL(Y_i,\bm{X}_i,\bm{\alpha})\right|\\
		\leq&\left\|\sum_{i=n+1}^{n+N}\bm{Z}_i^{\top}
		\left(\frac{1}{n}\sum_{i=1}^{n}\bm{Z}_i\bm{Z}_i^{\top}\right)^{-1}
		-NE(\bm{Z}^{\top})\{E(\bm{Z}\bm{Z}^{\top})\}^{-1}\right\|_2
		\left\|\frac{1}{n}\sum_{i=1}^{n}\bm{Z}_iL(Y_i,\bm{X}_i,\bm{\alpha})\right\|_2\\
		\leq&c_1\left\|\left\{\sum_{i=n+1}^{n+N}\bm{Z}_i^{\top}-NE(\bm{Z}^{\top})\right\}
		\left(\sum_{i=1}^{n}\bm{Z}_i\bm{Z}_i^{\top}\right)^{-1}\right\|_2\\
		&+c_1N\left\|E(\bm{Z}^{\top})
		\left[\left(\frac{1}{n}\sum_{i=1}^{n}\bm{Z}_i\bm{Z}_i^{\top}\right)^{-1}
		-E(\bm{Z}^{\top})\{E(\bm{Z}\bm{Z}^{\top})\}^{-1}\right]\right\|_2\\
		\leq&c_2(N^{1/2}+Nn^{-1/2}),
	\end{split}
\end{equation}
where $\|\cdot\|_2$ is $L_2$ norm. Then, by equations (A.1) and (A.2), and $n\rightarrow\infty$, we have
\begin{equation}
	\begin{split}
		&\sup_{\bm{\alpha}\in\Theta}|\tilde{L}(\bm{\alpha})/(n+N)
		-E\{L(\bm{Y},\bm{X},\bm{\alpha})\}|\\
		&\leq \sup_{\bm{\alpha}\in\Theta}|\tilde{L}(\bm{\alpha})-\bar{L}(\bm{\alpha})|/(n+N)
		+\sup_{\bm{\alpha}\in\Theta}|\bar{L}(\bm{\alpha})/(n+N)-E\{L(\bm{Y},\bm{X},\bm{\alpha})\}|
		=o_p(1).
	\end{split}
\end{equation}
(A.3) implies that $\|\text{Vec}(\tilde{\bm{\alpha}}-\bm{\alpha}^*)\|_2=o_p(1)$ according to $\bm{\alpha}^*$ is the unique minimizer of $E\{L(\bm{Y},\bm{X},\bm{\alpha})\}$.  Thus, the consistency is proved.
\\
\par
{\bf To show asymptotic normality.} By equation (3.2) and Taylor's expansion of $\nabla_{\text{Vec}(\bm{\alpha})}\tilde{L}(\bm{\alpha})|_{\bm{\alpha}=\tilde{\bm{\alpha}}}$ at $\bm{\alpha}^*$ as
\begin{equation}
	\begin{split}
		\bm{0}=\nabla_{\text{Vec}(\bm{\alpha})}\tilde{L}(\bm{\alpha})|_{\bm{\alpha}=\tilde{\bm{\alpha}}}
		=\nabla_{\text{Vec}(\bm{\alpha})}\tilde{L}(\bm{\alpha})|_{\bm{\alpha}=\bm{\alpha}^*}+
		\nabla^2_{\text{Vec}(\bm{\alpha})}
		\tilde{L}(\bm{\alpha})|_{\bm{\alpha}=\bar{\bm{\alpha}}}\text{Vec}(\tilde{\bm{\alpha}}
		-\bm{\alpha}^*),
	\end{split}
\end{equation}
where $\bar{\bm{\alpha}}$ is between $\tilde{\bm{\alpha}}$ and $\bm{\alpha}^*$.
We first consider $\nabla_{\text{Vec}(\bm{\alpha})}\tilde{L}(\bm{\alpha})|_{\bm{\alpha}=\bm{\alpha}^*}$.
Denote $\tilde{\bm{Z}}_N=\sum_{i=n+1}^{n+N}\bm{Z}_i/N$, $\tilde{\bm{Z}}_n=\sum_{i=1}^{n}\bm{Z}_i/n$, $\hat{{\bm \Sigma}}_{\bm{Z}}=\sum_{i=1}^{n}\bm{Z}_i\bm{Z}_i^{\top}/n$ and
$\bm{U}_i=S_i(\bm{\alpha}^*)-\bm{A}^{\top}\bm{Z}_i$ with $S_i(\bm{\alpha}^*)=\nabla_{\text{Vec}(\bm{\alpha})}L(Y_i,\bm{X}_i,\bm{\alpha})|_{\bm{\alpha}=\bm{\alpha}^*}$. Thus, we can obtain
\begin{equation}
	\begin{split}
		\nabla_{\text{Vec}(\bm{\alpha})}\tilde{L}(\bm{\alpha})|_{\bm{\alpha}=\bm{\alpha}^*}
		=&nS(\bm{\alpha}^*)+\frac{N}{n}\sum_{i=1}^{n}S_i(\bm{\alpha}^*)\bm{Z}_i^{\top}
		\hat{{\bm \Sigma}}_{\bm{Z}}^{-1}\tilde{\bm{Z}}_N\\
		=&nS(\bm{\alpha}^*)+\frac{N}{n}
		\sum_{i=1}^{n}(\bm{A}^{\top}\bm{Z}_i+\bm{U}_i)\bm{Z}_i^{\top}\hat{{\bm \Sigma}}_{\bm{Z}}^{-1}\tilde{\bm{Z}}_N\\
		=&nS(\bm{\alpha}^*)+N\bm{A}^{\top}\left\{\tilde{\bm{Z}}_N-E(\bm{Z})\right\}
		+N\bm{A}^{\top}E(\bm{Z})\\
		&+\frac{N}{n}\sum_{i=1}^{n}\bm{U}_i\bm{Z}_i^{\top}
		\hat{{\bm \Sigma}}_{\bm{Z}}^{-1}\tilde{\bm{Z}}_n+\frac{N}{n}\sum_{i=1}^{n}\bm{U}_i\bm{Z}_i^{\top}
		\hat{{\bm \Sigma}}_{\bm{Z}}^{-1}
		\left\{\tilde{\bm{Z}}_N-\tilde{\bm{Z}}_n\right\}\\
		=&\left\{nS(\bm{\alpha}^*)+\frac{N}{n}\sum_{i=1}^{n}\bm{U}_i\right\}
		+N\bm{A}^{\top}\left\{\tilde{\bm{Z}}_N-E(\bm{Z})\right\}\\
		&+o_p(n^{-1/2}(n+N))\\
		=&\left\{(n+N)S(\bm{\alpha}^*)-N\bm{A}^{\top}\tilde{\bm{Z}}_n\right\}
		+N\bm{A}^{\top}\left\{\tilde{\bm{Z}}_N-E(\bm{Z})\right\}\\
		&+o_p(n^{-1/2}(n+N)),
	\end{split}
\end{equation}
where
the forth equality holds because of $\bm{A}^{\top}E(\bm{Z})=\bm{0}$ by the definition of $\bm{\alpha}^*$, $
\hat{{\bm \Sigma}}_{\bm{Z}}^{-1}\tilde{\bm{Z}}_n=(1,\bm{0}_{d-1}^{\top})^{\top}$ by Lemma 2 in \cite{r47}, and $$\frac{N}{n}\sum_{i=1}^{n}\bm{U}_i\bm{Z}_i^{\top}
\hat{{\bm \Sigma}}_{\bm{Z}}^{-1}
\left\{\tilde{\bm{Z}}_N-\tilde{\bm{Z}}_n\right\}=o_p(n^{-1/2}(n+N))$$ by proof similar to Theorem 1 in \cite{r47}.
Finally, we consider $\nabla^2_{\text{Vec}(\bm{\alpha})}\tilde{L}(\bm{\alpha})|_{\bm{\alpha}=\bar{\bm{\alpha}}}$ as
\begin{equation}
	\begin{split}
		&\nabla^2_{\text{Vec}(\bm{\alpha})}\tilde{L}(\bm{\alpha})|_{\bm{\alpha}=\bar{\bm{\alpha}}}\\
		=&
		\nabla^2_{\text{Vec}(\bm{\alpha})}\bar{L}(\bm{\alpha})|_{\bm{\alpha}=\bar{\bm{\alpha}}}
		+\left\{\nabla^2_{\text{Vec}(\bm{\alpha})}\tilde{L}(\bm{\alpha})|_{\bm{\alpha}=\bar{\bm{\alpha}}}
		-\nabla^2_{\text{Vec}(\bm{\alpha})}\bar{L}(\bm{\alpha})|_{\bm{\alpha}=\bar{\bm{\alpha}}} \right\}
		\\
		=&(n+N){\bf H}+\left\{\nabla^2_{\text{Vec}(\bm{\alpha})}\bar{L}(\bm{\alpha})|_{\bm{\alpha}
			=\bar{\bm{\alpha}}}-(n+N){\bf H}\right\}
		+O_p(N^{1/2}+Nn^{-1/2})\\
		=&(n+N)\{{\bf H}+o_p(1)\},
	\end{split}
\end{equation}
where the second equation is similar to (A.2) by conditions {\bf C2} and {\bf C4}, and the last equation is according to (A.1). Then, from (A.4)-(A.6), we have
\begin{equation}
	\begin{split}
		\text{Vec}(\tilde{\bm{\alpha}}-\bm{\alpha}^*)
		=-{\bf H}^{-1}\left[\left\{S(\bm{\alpha}^*)-\frac{N}{n+N}\bm{A}^{\top}\tilde{\bm{Z}}_n
		\right\}
		+\frac{N}{n+N}\bm{A}^{\top}\left\{\tilde{\bm{Z}}_N-E(\bm{Z})\right\}
		\right]+o_p(n^{-1/2}).
	\end{split}
\end{equation}
Therefore, we can prove the theorem.
\\
\par
{\bf Proof of Theorem 3.2.} From the $\tilde{\beta}_{\tau}=\tilde{\alpha}\int_0^1\bm{b}(\bar{\tau})\text{J}_{\tau}(\bar{\tau})d\bar{\tau}$ and (A.7), the theorem can be directly proven.

\section*{Reference }
  \bibliographystyle{elsarticle-harv}
 \bibliography{ref}

\begin{thebibliography}{32}
\expandafter\ifx\csname natexlab\endcsname\relax\def\natexlab#1{#1}\fi
\providecommand{\url}[1]{\texttt{#1}}
\providecommand{\href}[2]{#2}
\providecommand{\path}[1]{#1}
\providecommand{\DOIprefix}{doi:}
\providecommand{\ArXivprefix}{arXiv:}
\providecommand{\URLprefix}{URL: }
\providecommand{\Pubmedprefix}{pmid:}
\providecommand{\doi}[1]{\href{http://dx.doi.org/#1}{\path{#1}}}
\providecommand{\Pubmed}[1]{\href{pmid:#1}{\path{#1}}}
\providecommand{\bibinfo}[2]{#2}
\ifx\xfnm\relax \def\xfnm[#1]{\unskip,\space#1}\fi
\bibitem[{Cai and Guo(2020)}]{r62}
\bibinfo{author}{Cai, T.}, \bibinfo{author}{Guo, Z.}, \bibinfo{year}{2020}.
\newblock \bibinfo{title}{Semisupervised inference for explained variance in
  high dimensional linear regression and its applications}.
\newblock \bibinfo{journal}{Journal of the Royal Statistical Society: Series B
  (Statistical Methodology)} \bibinfo{volume}{82}, \bibinfo{pages}{391--419}.
\newblock \DOIprefix\doi{10.1111/rssb.12357}.
\bibitem[{Cannings(2021)}]{r65}
\bibinfo{author}{Cannings, T.}, \bibinfo{year}{2021}.
\newblock \bibinfo{title}{Random projections: Data perturbation for
  classification problems}.
\newblock \bibinfo{journal}{WIREs Computational Statistics}
  \bibinfo{volume}{13}, \bibinfo{pages}{e1499}.
\newblock \DOIprefix\doi{10.1002/wics.1499}.
\bibitem[{Chakrabortty and Cai(2018)}]{r64}
\bibinfo{author}{Chakrabortty, A.}, \bibinfo{author}{Cai, T.},
  \bibinfo{year}{2018}.
\newblock \bibinfo{title}{Efficient and adaptive linear regression in
  semi-supervised settings}.
\newblock \bibinfo{journal}{Annals of Statistics} \bibinfo{volume}{46},
  \bibinfo{pages}{1541--1572}.
\newblock \DOIprefix\doi{10.1214/17-AOS1594}.
\bibitem[{Chapelle et~al.(2010)Chapelle, Scholkopf and Zien}]{r54}
\bibinfo{author}{Chapelle, O.}, \bibinfo{author}{Scholkopf, B.},
  \bibinfo{author}{Zien, A.}, \bibinfo{year}{2010}.
\newblock \bibinfo{title}{Semi-supervised learning}.
\newblock \bibinfo{journal}{the MIT Press} .
\bibitem[{Chen et~al.(2024)Chen, Mao and Yang}]{r78}
\bibinfo{author}{Chen, H.}, \bibinfo{author}{Mao, T.}, \bibinfo{author}{Yang,
  F.}, \bibinfo{year}{2024}.
\newblock \bibinfo{title}{Estimation of the adjusted standard-deviatile for
  extreme risks}.
\newblock \bibinfo{journal}{Scandinavian Journal of Statistics}
  \bibinfo{volume}{51}, \bibinfo{pages}{643--671}.
\newblock \DOIprefix\doi{https://doi.org/10.1111/sjos.12693}.
\bibitem[{Chen et~al.(2023)Chen, Ma and Sun}]{r50}
\bibinfo{author}{Chen, Y.}, \bibinfo{author}{Ma, M.}, \bibinfo{author}{Sun,
  H.}, \bibinfo{year}{2023}.
\newblock \bibinfo{title}{Statistical inference for extreme extremile in
  heavy-tailed heteroscedastic regression model}.
\newblock \bibinfo{journal}{Insurance: Mathematics and Economics}
  \bibinfo{volume}{111}, \bibinfo{pages}{142--162}.
\newblock \DOIprefix\doi{10.1016/j.insmatheco.2023.04.001}.
\bibitem[{Cheplygina et~al.(2019)Cheplygina, de~Bruijne and Pluim}]{r56}
\bibinfo{author}{Cheplygina, V.}, \bibinfo{author}{de~Bruijne, M.},
  \bibinfo{author}{Pluim, J.}, \bibinfo{year}{2019}.
\newblock \bibinfo{title}{Not-so-supervised: A survey of semi-supervised,
  multi-instance, and transfer learning in medical image analysis}.
\newblock \bibinfo{journal}{Medical Image Analysis} \bibinfo{volume}{54},
  \bibinfo{pages}{280--296}.
\newblock \DOIprefix\doi{10.1016/j.media.2019.03.009}.
\bibitem[{Daouia et~al.(2019)Daouia, Gijbels and Stupfler}]{r1}
\bibinfo{author}{Daouia, A.}, \bibinfo{author}{Gijbels, I.},
  \bibinfo{author}{Stupfler, G.}, \bibinfo{year}{2019}.
\newblock \bibinfo{title}{Extremiles: A new perspective on asymmetric least
  squares}.
\newblock \bibinfo{journal}{Journal of the American Statistical Association}
  \bibinfo{volume}{114}, \bibinfo{pages}{1366--1381}.
\newblock \DOIprefix\doi{10.1080/01621459.2018.1498348}.
\bibitem[{Daouia et~al.(2022)Daouia, Gijbels and Stupfler}]{r41}
\bibinfo{author}{Daouia, A.}, \bibinfo{author}{Gijbels, I.},
  \bibinfo{author}{Stupfler, G.}, \bibinfo{year}{2022}.
\newblock \bibinfo{title}{Extremile regression}.
\newblock \bibinfo{journal}{Journal of the American Statistical Association}
  \bibinfo{volume}{117}, \bibinfo{pages}{1579--1586}.
\newblock \DOIprefix\doi{10.1080/01621459.2021.1875837}.
\bibitem[{Flutre et~al.(2013)Flutre, Wen, Pritchard and Stephens}]{r61}
\bibinfo{author}{Flutre, T.}, \bibinfo{author}{Wen, X.},
  \bibinfo{author}{Pritchard, J.}, \bibinfo{author}{Stephens, M.},
  \bibinfo{year}{2013}.
\newblock \bibinfo{title}{A statistical framework for joint eqtl analysis in
  multiple tissues}.
\newblock \bibinfo{journal}{PLoS genetics} \bibinfo{volume}{9},
  \bibinfo{pages}{e1003486}.
\newblock \DOIprefix\doi{10.1371/journal.pgen.1003486}.
\bibitem[{Frumento and Bottai(2016)}]{r44}
\bibinfo{author}{Frumento, P.}, \bibinfo{author}{Bottai, M.},
  \bibinfo{year}{2016}.
\newblock \bibinfo{title}{Parametric modeling of quantile regression
  coefficient functions}.
\newblock \bibinfo{journal}{Biometrics} \bibinfo{volume}{72},
  \bibinfo{pages}{74--84}.
\newblock \DOIprefix\doi{10.1111/biom.12410}.
\bibitem[{Furno(2023)}]{r80}
\bibinfo{author}{Furno, M.}, \bibinfo{year}{2023}.
\newblock \bibinfo{title}{Extremiles, quantiles and expectiles in the tails}.
\newblock \bibinfo{journal}{Journal of Computational Finance}
  \bibinfo{volume}{27}, \bibinfo{pages}{87--113}.
\newblock \DOIprefix\doi{10.21314/JCF.2023.011}.
\bibitem[{Geng(2024)}]{r81}
\bibinfo{author}{Geng, Z.}, \bibinfo{year}{2024}.
\newblock \bibinfo{title}{Modelling additive extremile regression by
  iteratively penalized least asymmetric weighted squares and gradient descent
  boosting}.
\newblock \bibinfo{journal}{Statistics} \bibinfo{volume}{58},
  \bibinfo{pages}{576--595}.
\newblock \DOIprefix\doi{10.1080/02331888.2024.2348077}.
\bibitem[{Jiang et~al.(2024)Jiang, Choy and Yu}]{r77}
\bibinfo{author}{Jiang, R.}, \bibinfo{author}{Choy, S.K.}, \bibinfo{author}{Yu,
  K.}, \bibinfo{year}{2024}.
\newblock \bibinfo{title}{Non-crossing quantile double-autoregression for the
  analysis of streaming time series data}.
\newblock \bibinfo{journal}{Journal of Time Series Analysis}
  \bibinfo{volume}{45}, \bibinfo{pages}{513--532}.
\newblock \DOIprefix\doi{10.1111/jtsa.12725}.
\bibitem[{Jiang and Yu(2023)}]{r66}
\bibinfo{author}{Jiang, R.}, \bibinfo{author}{Yu, K.}, \bibinfo{year}{2023}.
\newblock \bibinfo{title}{No-crossing single-index quantile regression curve
  estimation}.
\newblock \bibinfo{journal}{Journal of Business \& Economic Statistics}
  \bibinfo{volume}{41}, \bibinfo{pages}{309--320}.
\newblock \DOIprefix\doi{10.1080/07350015.2021.2013245}.
\bibitem[{Jiang and Yu(2024)}]{r76}
\bibinfo{author}{Jiang, R.}, \bibinfo{author}{Yu, K.}, \bibinfo{year}{2024}.
\newblock \bibinfo{title}{Unconditional quantile regression for streaming data
  sets}.
\newblock \bibinfo{journal}{Journal of Business \& Economic Statistics}
  \bibinfo{volume}{42}, \bibinfo{pages}{1143--1154}.
\newblock \DOIprefix\doi{10.1080/07350015.2003.2293162}.
\bibitem[{Kriegler and Berk(2010)}]{r52}
\bibinfo{author}{Kriegler, B.}, \bibinfo{author}{Berk, R.},
  \bibinfo{year}{2010}.
\newblock \bibinfo{title}{Small area estimation of the homeless in los angeles:
  an application of cost-sensitive stochastic gradient boosting}.
\newblock \bibinfo{journal}{Annals of Applied Statistics} \bibinfo{volume}{4},
  \bibinfo{pages}{1234--1255}.
\newblock \DOIprefix\doi{10.1214/10-AOAS328}.
\bibitem[{Leblanc et~al.(2006)Leblanc, Moon and Kooperberg}]{r71}
\bibinfo{author}{Leblanc, M.}, \bibinfo{author}{Moon, J.},
  \bibinfo{author}{Kooperberg, C.}, \bibinfo{year}{2006}.
\newblock \bibinfo{title}{Extreme regression}.
\newblock \bibinfo{journal}{Biostatistics} \bibinfo{volume}{7},
  \bibinfo{pages}{71--84}.
\newblock \DOIprefix\doi{10.1093/biostatistics/kxi041}.
\bibitem[{Mcleish and Tosh(1983)}]{r74}
\bibinfo{author}{Mcleish, D.}, \bibinfo{author}{Tosh, D.},
  \bibinfo{year}{1983}.
\newblock \bibinfo{title}{The estimation of extreme quantiles in logit
  bioassay}.
\newblock \bibinfo{journal}{Biometrika} \bibinfo{volume}{70},
  \bibinfo{pages}{625--632}.
\newblock \DOIprefix\doi{10.1093/biomet/70.3.625}.
\bibitem[{Mendoza and De~la Hoz~Manotas(2019)}]{r73}
\bibinfo{author}{Mendoza, F.}, \bibinfo{author}{De~la Hoz~Manotas, A.},
  \bibinfo{year}{2019}.
\newblock \bibinfo{title}{Dataset for estimation of obesity levels based on
  eating habits and physical condition in individuals from colombia, peru and
  mexico}.
\newblock \bibinfo{journal}{Data in Brief} \bibinfo{volume}{25},
  \bibinfo{pages}{104344}.
\newblock \DOIprefix\doi{10.1016/j.dib.2019.104344}.
\bibitem[{Michaelson et~al.(2009)Michaelson, Loguercio and Beyer}]{r59}
\bibinfo{author}{Michaelson, J.}, \bibinfo{author}{Loguercio, S.},
  \bibinfo{author}{Beyer, A.}, \bibinfo{year}{2009}.
\newblock \bibinfo{title}{Detection and interpretation of expression
  quantitative trait loci (eqtl)}.
\newblock \bibinfo{journal}{Methods (San Diego, Calif.)} \bibinfo{volume}{48},
  \bibinfo{pages}{265--76}.
\newblock \DOIprefix\doi{10.1016/j.ymeth.2009.03.004}.
\bibitem[{Newey and Powell(1987)}]{r70}
\bibinfo{author}{Newey, W.K.}, \bibinfo{author}{Powell, J.L.},
  \bibinfo{year}{1987}.
\newblock \bibinfo{title}{Asymmetric least squares estimation and testing}.
\newblock \bibinfo{journal}{Econometrica} \bibinfo{volume}{55},
  \bibinfo{pages}{819--847}.
\bibitem[{Song et~al.(2024a)Song, Lin and Zhou}]{r47}
\bibinfo{author}{Song, S.}, \bibinfo{author}{Lin, Y.}, \bibinfo{author}{Zhou,
  Y.}, \bibinfo{year}{2024}a.
\newblock \bibinfo{title}{A general m-estimation theory in semi-supervised
  framework}.
\newblock \bibinfo{journal}{Journal of the American Statistical Association}
  \bibinfo{volume}{119}, \bibinfo{pages}{1065--1075}.
\newblock \DOIprefix\doi{10.1080/01621459.2023.2169699}.
\bibitem[{Song et~al.(2024b)Song, Lin and Zhou}]{r83}
\bibinfo{author}{Song, S.}, \bibinfo{author}{Lin, Y.}, \bibinfo{author}{Zhou,
  Y.}, \bibinfo{year}{2024}b.
\newblock \bibinfo{title}{Semi-supervised inference for block-wise missing data
  without imputation}.
\newblock \bibinfo{journal}{Journal of Machine Learning Research}
  \bibinfo{volume}{25}, \bibinfo{pages}{1--36}.
\newblock \URLprefix \url{http://jmlr.org/papers/v25/21-1504.html}.
\bibitem[{Sun and Wang(2024)}]{r82}
\bibinfo{author}{Sun, W.}, \bibinfo{author}{Wang, S.}, \bibinfo{year}{2024}.
\newblock \bibinfo{title}{Remire: Robust extremile regression in high
  dimensions}.
\newblock \bibinfo{journal}{Doi: 10.21203/rs.3.rs-5161987/v1} .
\bibitem[{Tauchen(1985)}]{r48}
\bibinfo{author}{Tauchen, G.}, \bibinfo{year}{1985}.
\newblock \bibinfo{title}{Diagnostic testing and evaluation of maximum
  likelihood models}.
\newblock \bibinfo{journal}{Journal of Econometrics} \bibinfo{volume}{30},
  \bibinfo{pages}{415--443}.
\newblock \DOIprefix\doi{10.1016/0304-4076(85)90149-6}.
\bibitem[{Ullah et~al.(2023)Ullah, Wang and Yao}]{r69}
\bibinfo{author}{Ullah, A.}, \bibinfo{author}{Wang, T.}, \bibinfo{author}{Yao,
  W.}, \bibinfo{year}{2023}.
\newblock \bibinfo{title}{Semiparametric partially linear varying coefficient
  modal regression}.
\newblock \bibinfo{journal}{Journal of Econometrics} \bibinfo{volume}{235},
  \bibinfo{pages}{1001--1026}.
\newblock \DOIprefix\doi{10.1016/j.jeconom.2022.09.002}.
\bibitem[{Wang et~al.(2019)Wang, Lin, Cui, Jia, Wang, Fang, Yu, Zhou, Yang and
  Qi}]{r57}
\bibinfo{author}{Wang, D.}, \bibinfo{author}{Lin, J.}, \bibinfo{author}{Cui,
  P.}, \bibinfo{author}{Jia, Q.}, \bibinfo{author}{Wang, Z.},
  \bibinfo{author}{Fang, Y.}, \bibinfo{author}{Yu, Q.}, \bibinfo{author}{Zhou,
  J.}, \bibinfo{author}{Yang, S.}, \bibinfo{author}{Qi, Y.},
  \bibinfo{year}{2019}.
\newblock \bibinfo{title}{A semi-supervised graph attentive network for nancial
  fraud detection}.
\newblock \bibinfo{journal}{2019 IEEE International Conference on Data Mining}
  , \bibinfo{pages}{598--607}.
\bibitem[{Wang and Lian(2020)}]{r68}
\bibinfo{author}{Wang, L.}, \bibinfo{author}{Lian, H.}, \bibinfo{year}{2020}.
\newblock \bibinfo{title}{Communication-efficient estimation of
  high-dimensional quantile regression}.
\newblock \bibinfo{journal}{Analysis and Applications} \bibinfo{volume}{18},
  \bibinfo{pages}{1057--1075}.
\newblock \DOIprefix\doi{10.1142/S0219530520500098}.
\bibitem[{Wen et~al.(2024)Wen, Jia, Ren, Wang and Zou}]{r79}
\bibinfo{author}{Wen, M.}, \bibinfo{author}{Jia, Y.}, \bibinfo{author}{Ren,
  H.}, \bibinfo{author}{Wang, Z.}, \bibinfo{author}{Zou, C.},
  \bibinfo{year}{2024}.
\newblock \bibinfo{title}{Semi-supervised distribution learning}.
\newblock \bibinfo{journal}{Biometrika} \DOIprefix\doi{10.1093/biomet/asae056}.
\bibitem[{Yuval and Rosset(2022)}]{r55}
\bibinfo{author}{Yuval, O.}, \bibinfo{author}{Rosset, S.},
  \bibinfo{year}{2022}.
\newblock \bibinfo{title}{Semi-supervised empirical risk minimization: Using
  unlabeled data to improve prediction}.
\newblock \bibinfo{journal}{Electronic Journal of Statistics}
  \bibinfo{volume}{16}, \bibinfo{pages}{1434--1460}.
\bibitem[{Zhang et~al.(2019)Zhang, Brown and Cai}]{r58}
\bibinfo{author}{Zhang, A.}, \bibinfo{author}{Brown, L.}, \bibinfo{author}{Cai,
  T.}, \bibinfo{year}{2019}.
\newblock \bibinfo{title}{Semi-supervised inference: General theory and
  estimation of means}.
\newblock \bibinfo{journal}{Annals of Statistics} \bibinfo{volume}{47},
  \bibinfo{pages}{2538--2566}.
\newblock \DOIprefix\doi{10.1214/18-AOS1756}.

\end{thebibliography}
 \end{document}